\documentclass[preprint2]{aastex}

\usepackage{epsfig}
\shortauthors{Robinson, Ivans, and Welsh}
\shorttitle{The Spectrum of CI~Cam/RXTE~J0421+560}

\begin{document}

\title{High-Dispersion Spectroscopy of the X-Ray Transient RXTE~J0421+560 
(= CI~Cam) during Outburst\footnote{Based on observations made with
the NASA/ESA Hubble Space Telescope, otained at the Space Telescope
Science Institute, which is operated by the Association of Universities
for Research in Astronomy, Inc., under NASA contract NAS 5-26555.}
}

\author{Edward L. Robinson, Inese I. Ivans, and William F. Welsh\footnote{
Current address: Department of Astronomy, San Diego State
University, 5500 Campanile Dr., San Diego, CA 92182}}
\affil{Department of Astronomy, The University of Texas,
    Austin, TX 78712}

\begin{abstract}
We obtained high dispersion spectroscopy of CI~Cam, the optical
counterpart of XTE~J0421+560, two weeks after the peak of
its short outburst in 1998 April.  
The optical counterpart is a supergiant B[e] star that is emitting
a two-component wind, a cool, low-velocity wind
and a hot, high-velocity wind.
The cool wind, which is the source of narrow emission lines of
neutral and ionized metals, 
has a velocity of 32~km~s$^{-1}$ and a
temperature near $8000\ K$.
It is dense, roughly spherical, fills the space around the sgB[e] star,
and, based on the size of an infrared-emitting dust shell around the 
system, extends to a radius between 13~AU and 50 AU.
It carries away mass at a high rate, 
$\dot M > 10^{-6}\ M_\odot\ \textrm{yr}^{-1}$.
The hot wind has a velocity in excess of 2500~km~s$^{-1}$
and a temperature of $1.7\pm0.3 \times 10^4~K$.
From an ultraviolet spectrogram of CI~Cam obtained in 2000 March
with Hubble Space Telescope, we derive a differential extinction
$E(B-V) = 0.85 \pm 0.05$. 

We show that the distance to CI~Cam is greater than 5~kpc.
Based on this revised distance, the X-ray luminosity at the peak 
of the outburst was
$L(2-25\ \rm{keV})\ > \ 3.0 \times 10^{38}\ \rm{erg\ s}^{-1}$,
making CI~Cam one of the most luminous X-ray transients.
The ratio of quiescent luminosity to peak luminosity in the 2 -- 25~keV
band is $L_q/L_p < 1.7 \times 10^{-6}$.

The compact star in CI~Cam is immersed in the dense
circumstellar wind from the sgB[e] star and
burrows through the wind producing little X-ray emission
except for rare transient outbursts.
This picture, a compact star traveling in a wide orbit
through the dense circumstellar envelope of a sgB[e] star,
occasionally producing transient X-ray outbursts, makes
CI~Cam unique among the known X-ray binaries.
There is strong circumstantial evidence that the compact object 
is a black hole, not a neutron star.
We speculate that the X-ray outburst was short because
the accretion disk around the compact star is fed from a
stellar wind and is smaller than disks fed by Roche-lobe overflow.

\end{abstract}


\section{\bf I. Introduction}

The outburst of the transient X-ray source XTE~J0421+560 began 
1998 March 31.6 UTC and peaked only $\sim 12$ hours later
at nearly 1.9~crab (2 - 10 keV).
The outburst faded rapidly, initially on an
e-folding time scale of 0.6 days and then slowing to
a time scale of 2.3 days, reaching quiescence in less than
two weeks \citep{Bell99}.
There were no coherent pulsations in the X-ray light
curve with amplitudes greater than 0.1\% rms at
frequencies between 0.01 and 4096 Hz,
nor quasi-periodic oscillations with amplitudes 
greater than 0.7\%.
The outburst was also detected by the PCA on RXTE, and by
BATSE, BeppoSAX, and ASCA at energies from 0.3 to 75 keV
\citep{Fron98, Harm98, Paci98, Ueda98, Bell99}.
The flux distribution at energies greater than 1~keV was consistent
with either a two-temperature thermal bremsstrahlung spectrum (1.1 and
5.7 keV) or a power law with a high-energy cutoff;
the spectrum softened as CI~Cam faded towards quiescence.
In addition, there was an extremely soft component confined 
to energies less than 1~keV that appeared only during two 
short flares \citep{Ueda98}.
Weak X-ray emission was also detected in quiescence, 157 days after
the outburst \citep{Orla00}.

The optical counterpart to J0421+560 is CI~Cam (= MWC~84 = MW~143) 
\citep{Smit98, Hjel98a, Wagn98, Robi98}.
Before its outburst CI~Cam was a low-amplitude, irregular 
variable star with $<\! V\! >\ = 11.6$.
There was no convincing pattern to its variations \citep{Berg95},
although \citet{Miro95} has suggested that the variations 
showed an 11.7~d quasi-period.
The pre-outburst spectrum of CI~Cam had strong, broad emission
lines of H and He~I, and narrow emission lines of Fe~II, 
all superimposed on a hot, blue continuum;
there were no absorption lines, nor were there emission lines from
highly ionized species such as He~II and [O~III] (Downes 1984).
The spectrum was not noticeably different in spectrograms taken as
early as 1931 \citep{Merr33}.
\citet{Alle73} found a strong infrared excess,
which he attributed to thermal emission from circumstellar
dust with temperatures between 1190 and $1350~K$.
According to \citet{Zore98}, the distance to CI~Cam is
1750~pc and its luminosity is $M_{bol} = -6.9$.

These properties clearly place CI~Cam among the supergiant B[e] stars 
\citep{Zick98, Miro98, Lame98}.
The sgB[e] stars are massive, evolved, high-luminosity 
stars undergoing mass loss in a two-component wind.
The rates of mass loss are prodigious, 
typically greater than $10^{-5}\ M_\odot\ \textrm{yr}^{-1}$
\citep{Pach98}.
The notation ``B[e]'' or ``sgB[e]" risks confusing these stars with
the ordinary Be stars, which are
rapidly rotating stars near the main sequence losing mass in
an equatorial wind.
In practice, however, the spectroscopic and photometric properties 
of the sgB[e] stars are easily distinguished from those of
Be stars.

The optical, infrared, and radio light curves of the 1998 outburst
have been collected together by \citet{Clar00}.
CI~Cam rose by at least 3.4 magnitudes in the $R$ band, reaching
$R \approx 7.1$, and then faded towards quiescence, initially
on an e-folding time scale of $3.4 \pm 0.4$~days and then slowing 
to an e-folding time of $\sim 24$ days a few weeks after the 
outburst peak. 
Although the optical flux soon returned to its quiescent level,
the near-infrared flux remained $\sim 0.5$ magnitudes
above the pre-outburst flux for at least one year after the outburst,
which led \citet{Clar00} to infer that either the structure or 
composition of the dust shell had changed.
The radio light curves  were consistent with synchrotron emission
from an expanding cloud.
VLA maps suggested that CI~Cam ejected relativistic corkscrew jets 
during the outburst but the jets have not been
confirmed by other observations \citep{Hjel98b}.

The equivalent widths of the emission lines in the optical spectrum
of CI~Cam increased substantially during the outburst.
The He~I line at 6876~\AA\ reached almost 400~\AA\ before dropping back to
$\sim$50~\AA\ in quiescence, and the 4686~\AA\ line of He~II appeared 
in emission and grew to an equivalent width of $\sim 70$~\AA\ before fading
to $<2$~\AA\ 40 days later \citep{Bars98}.
The Na~I D lines, invisible in quiescence, reached an equivalent 
width near 40~\AA.
By early 1999, 10 months after the outburst, 
the optical spectrum was again similar to the pre-outburst 
spectrum \citep{Orla00}.

The outburst of J0421+560/CI~Cam was unusually short for an
X-ray transient and, accepting for the moment the distance given
by \citet{Zore98}, its luminosity was rather low, a 
few times $10^{37}\ \textrm{ergs s}^{-1}$.
This suggested to \citet{Bell99} that
CI~Cam is similar to A0538-66, an X-ray binary containing an
ordinary Be star and a neutron star.
The neutron star in A0538-66 has a rotation period of 0.067~s,
an orbital period of 16.65 days, and an orbital eccentricity
greater than 0.8 \citep{Bild97}.
The system has transient X-ray outbursts that recur at
multiples of the orbital period, presumably
when the neutron star passes through the equatorial wind from
the Be star.
The X-ray outbursts have durations between a few hours and
10 days and luminosities from $\sim 10^{37}\ 
\textrm{to}\ 10^{39}\ \textrm{erg s}^{-1}$, but the
outbursts can disappear altogether for many years,
possibly because the equatorial wind is strongly variable
\citep{Corb97}.

In this paper we present high-dispersion, optical spectroscopy of CI~Cam
obtained two weeks after the peak of the 1998 outburst, 
and ultraviolet spectroscopy obtained with HST in 2000 March.
We use the data to investigate the structure of the CI~Cam system
and elucidate the nature of the X-ray transient.

\section{Observations}

\subsection{Optical Spectroscopy and Line Identifications}

We measured the spectrum of CI~Cam on 1998 April 14, 15 and 17 UTC
with the coud\'e echelle spectrograph of the 2.7-m telescope at
McDonald Observatory \citep{Tull95}.
The spectrograms have a resolution of R = 60,000 and a useful
spectral range from $\sim$4300~\AA\ to 1.02 $\mu$m
except for inter-order gaps at longer wavelengths.
We observed CI~Cam for
$\sim$1.5 hours each night before it set in the west, obtaining
three or four individual exposures of 15 -- 30 minutes each night.
Although these spectra were obtained only two weeks after the
beginning of the outburst,
CI~Cam had already faded to within 0.5 mag of minimum light and
its brightness was no longer changing rapidly.

There are no absorption lines in the visual spectrum of CI~Cam
other than interstellar absorption features 
but, in compensation, CI~Cam has a rich emission-line spectrum.
Some of the lines are extraordinarily strong:
H$\alpha$ had an equivalent width of $\sim$750~\AA, the 
O~I line at 8446~\AA\ had an equivalent width of $\sim$175~\AA,
and the He~I triplet line at 7065~\AA\ had an equivalent width
of $\sim$145~\AA.
The list of species for which we have certain identifications
is given in Table~1.
The identifications are conservative and were accepted only if
supported by several emission lines.

None of the emission lines showed detectable radial velocity
variations during the observations.  
The upper limit to the variations is
\( \pm 0.5\ {\rm km\ s^{-1}} \) except for the He~II lines, for
which the upper limit is \( \pm 1.5\ {\rm km\ s^{-1}} \).
The He~II line at 4686~\AA\ is contaminated by an
instrumental artifact and the He~II line at 10123~\AA\ line 
is contaminated by OH emission from the night sky \citep{Oste97}, 
so we exclude He~II from further discussion.

The heliocentric radial velocities of individual species run from
\( -42.6\ {\rm km~s^{-1}} \) for [N~II]
to \( -46.4\ {\rm km~s^{-1}} \) for the Paschen lines, but
we consider the differences among these velocities to be 
the upper limit to any real differences because the profiles 
of the emission lines vary from species to species.
The mean radial velocity for all species we measured is
\( -44.4\pm0.6\ {\rm km~s^{-1}} \), where the standard deviation
refers to statistical errors;  systematic errors introduced by the
asymmetric line profiles could be several times larger.
The radial velocity with respect to the Local Standard of Rest
is  \( -51\ {\rm km~s^{-1}} \).

Figure~1 shows the region of the spectrum near 5860~\AA.
The upper panel shows the strongly asymmetric He~I line at 5876~\AA\ 
and the Na~D lines at 5890/96~\AA\ (dissected by narrow
interstellar absorption lines),
and the lower panel shows the same portion of the spectrum magnified
vertically to show the wings of the helium line and also
the weaker lines in the spectrum.
The blue wing of the He~I line extends to at least 5845~\AA\ 
($-1500\ \textrm{km~s}^{-1}$); the maximum extent of the red wing
is uncertain because of the overlapping Na~D lines.
The weak absorption feature at 5850~\AA\ is a diffuse
interstellar band \citep{Krel97}.

The thicket of weak emission lines seen in the blue wing of 
He~I 5876~\AA\ extends throughout the optical spectrum.
Figure~2 shows the region of the spectrum near 5160~\AA\
and provides identifications for the stronger lines.
Most of the lines come from Fe~II
but many come from Ti~II and Mg~I.
Forbidden Fe~II lines are present throughout the spectrum
(e.g., $\lambda\lambda$ 5158.00/5158.81~\AA) but they are generally
much weaker than the permitted Fe~II lines.

Figure~3 shows the portion of the infrared spectrum of CI~Cam
containing the O~I line at 8446~\AA, some members of the hydrogen Paschen 
series, some Fe~I lines, and
the Ca~II 8498~\AA\ line --  a member of the Ca~II infrared triplet.
The Paschen lines can be traced to P34 in other orders of
the spectrogram.
The O~I line at 8446~\AA\ is particularly interesting because it 
can be enhanced by  resonance-fluorescence 
\citep{Gran75, Kast95}.
The strongest transition of the 
\({\rm 2p^4 \phantom{\,}^3P - 3d \phantom{\,}^3D^0}\)
multiplet (UV 4) of O~I has a wavelength of 1025.77~\AA.\ 
O~I atoms in the ground state can absorb 
Ly$\beta$ photons at 1025.72~\AA\ and then
electrons in the \( {\rm 3d \phantom{\,}^3D^\circ} \) level can
branch to the \( {\rm 3p \phantom{\,}^3P} \) level,
emitting a photon at 11287~\AA\  followed by one at 8446~\AA.

Figure~4 shows another portion of the infrared
spectrum, with the N~I lines from multiplets 1 and 8, 
some Paschen lines, more Fe~I and [Fe~II] lines, and
Ca~II 8662~\AA\ -- another member of the infrared triplet.

\subsection{Ultraviolet Observations and Reddening}

We derived the reddening of CI~Cam from an ultraviolet
spectrogram obtained with the
Space Telescope Imaging Spectrograph on the Hubble Space Telescope
on 20 March 2000, two years after the outburst.
A more complete discussion of spectrum will
be given elsewhere; here we use the spectrum only to extract
the ultraviolet and optical extinction.

The spectrum was observed with the MAMA detector
and the medium-resolution echelle gratings E140M in the far
ultraviolet (1150~\AA\ -- 1730~\AA) and E230M in the near
ultraviolet (1870~\AA\ -- 2700~\AA).
To define the continuum distribution more accurately we
rebinned the two spectrograms into 15~\AA\ bins in
the short wavelength region and 20\ \AA\  bins in
the long wavelength region.
The rebinned spectrograms are shown in the lower panel of Figure~5;
the deep, broad dip at 2200~\AA\ shows that
the spectrum of CI~Cam is heavily reddened.

We adopted the extinction law given by \citet{Card89},
which has two parameters: $A_V$ 
and $R_V\ [= A_V/E(B-V)]$, where $A_V$ is the visual extinction
and $E(B-V)$ is the differential extinction 
between the $B$ and $V$ bands.
We de-extincted the spectrum of CI~Cam by requiring the de-extincted
spectrum to match the spectral distribution of
the HST standard star BD$+33^\circ2642$ between 1250~\AA\ and 
2650~\AA.
We chose BD$+33^\circ2642$ because it is a hot subgiant 
(spectral type B3~IV) and is likely to have a
continuum spectral distribution similar to that of CI~Cam.
The reddening of BD$+33^\circ2642$ is low but not zero:
According to \citet{Tobi85} its reddening is
$E(B-V) \sim E(b-y) = 0.011$, while \citet{Napi93} finds
$E(B-V) = 0.06$.
We have, therefore, applied a correction of
0.03 to the value of $E(B-V)$ derived from the fit, and a
correction of 0.1 to $A_V$.
We do not attempt to separate circumstellar from
interstellar extinction.

The results of the fit, including the after-the-fit corrections to
$E(B-V)$ and $A_V$, are
\begin{eqnarray}
     A_V & = & 2.3 \pm 0.3  \nonumber \\
     R_V  & = & 2.7 \pm 0.2 \nonumber \\
     E(B-V) & = & 0.85 \pm 0.05 \label{extinct1}
\end{eqnarray}
The quoted errors include the 
uncertainty in the reddening for BD$+33^\circ2642$.
The values of $A_V$ and $R_V$ derived from the fit are strongly
correlated, which increases their estimated error significantly.
The de-extincted spectrum of CI~Cam is shown in the upper panel
of Figure~5, with the
spectrum of BD$+33^\circ2642$ overplotted for comparison.
The fit of the de-extincted spectrum of CI~Cam to the spectrum
of BD$+33^\circ2642$ is quite good.
The high points between
2300~\AA\  and 2500~\AA, the high point near 1900~\AA, and the various dips
between 1200~\AA\ and 1700~\AA\ are all due to real features in the CI Cam
spectrum: Fe~II emission between 2300~\AA\ and 2500~\AA\ and wind features
at the shorter wavelengths, for example.
The value of $R_V$ derived from our data is significantly lower than the
commonly accepted value of 3.1, but 
the variance of the fit increases by more than a factor of two
if $R_V$ is increased from 2.7 to 3.1 and the fit looks markedly worse to
the eye.
The derived reddening does depend on the use of
BD$+33^\circ2642$ as a reddening template, but the dependence is
not strong.
Furthermore, $R_V = 2.7$ is well within the range of
measured values for individual stars (Cardelli et al.\ 1989).

\citet{Bell99} estimated the extinction and reddening
of CI~Cam in quiescence from the slope of the
optical/infrared continuum, finding
$A_V = 4.4 \pm 0.2,\ R_V = 3.7 \pm 0.1,\ \rm{and}\ E(B-V) = 1.18 \pm 0.04$.
The estimates by \citet{Zore98} agree fairly well with these numbers.
We attribute the difference between their values and ours 
to the substantial difficulties in deriving an accurate reddening 
from the slope of the optical continuum in an unusual star like
CI~Cam.
\citet{Clar00} derived $E(B-V) = 0.65 \pm 0.2$ from the 
strength of the diffuse interstellar bands in the optical spectrum, 
in rough agreement with our results.
The extinction at soft X-ray wavelengths yielded
$N_H = 3.76 \pm 0.36 \times 10^{22}\ \rm{cm}^{-2}$ near the
peak of the X-ray outburst, but $N_H$ rapidly decreased to
less than about $0.2 \times 10^{22}\ \rm{cm}^{-2}$ as CI~Cam
approached quiescence \citet{Bell99}.
If we were to adopt
$N_H/A_V = 1.79 \times 10^{21}\ \rm{atoms\ cm}^{-2}\ \rm{mag}^{-1}$,
a typical value for the interstellar medium
(e.g., Predehl \& Schmitt 1995), the visual extinction would be
only 1.1 mag during quiescence.
The ratio of optical to X-ray extinction is, therefore, higher
than normal.

The change in the infrared flux after the outburst and 
the rapid change of the soft X-ray extinction during the
outburst \citep{Clar00, Bell99},
demonstrate that much of the extinction to
CI~Cam is local, not interstellar, raising a concern
that the reddening measured from the ultraviolet
spectrum in 2000 March may not be the same as
the visual reddening in 1998 April.
Because of this concern, most of the calculations we present later
in this paper will include a much wider range of values
for $E(B-V)$ than permitted by the standard deviations 
given in equation~\ref{extinct1}.

\subsection{The Distance to CI~Cam}

Although the spectrum of CI~Cam clearly places it among the sgB[e] stars, 
the sgB[e] stars are a diverse group with a wide range of luminosities
\citep{Zick98}.
\citet{Zore98}, \citet{Bell99}, \citet{Clar00}
and \citet{Orla00} have all argued that CI~Cam
is located at a distance of $\sim2$ ~kpc
and has a luminosity in the range $10^{4.7} - 10^{4.9}\ L_\odot$. 
We argue here that CI~Cam is more distant and more luminous
than these estimates.

\citet{Clar00} and others estimated the
distance to CI~Cam from its optical continuum extinction and 
from the strength of the diffuse interstellar bands.
The extinction can be converted to distance using, for
example, the extinction/distance graphs in \citet{Neck80}.
The extinction in the direction of CI~Cam is, however, patchy
and sparsely measured, so the extinction/distance relation is
not well determined.
In fact, the line of sight to CI~Cam passes through a
relatively clear window between molecular clouds, making any
mean extinction/distance relation inapplicable
\citep{Dame87}.
Also, CI~Cam lies at a galactic latitude of $4\fdg1$, well
above the midplane of the galaxy as defined by the neutral
hydrogen layer in that direction \citep{Kerr86}.
The line of sight to CI~Cam rises to 150~pc above the plane of
the galaxy at a distance of $\sim$2~kpc.
Beyond this distance the line of sight exits the densest absorbing
layer and encounters little further extinction until, at
distances greater than $\sim$ 5 kpc, the line of sight begins to
re-enter the greatly-warped outer disk of the galaxy.
Interstellar extinction gives, then, only a lower limit for the
distance to CI~Cam -- and not a particularly reliable one.

\citet{Bell99} show that CI~Cam must be more distant
than 350~pc and then simply assign CI~Cam to the
Perseus arm, which is at a distance of $\sim$2~kpc.
H~II regions and young clusters are not, however, strongly confined
to the Perseus arm in the direction of CI~Cam and some are
found many kiloparsecs beyond the Perseus arm \citep{Tayl93}, so the
argument for placing CI~Cam in the Perseus arm is not strong.
The distance determinations by \citet{Zore98} rely on uncertain
measurements and relationships, and could easily
be in error by large factors for individual stars.

Two lines of evidence show that CI~Cam is much more distant than 2~kpc.
The first is based on the spectrum of CI~Cam.
Although helium lines are present in the spectra of most
sgB[e] stars, the lines are often present in absorption, not emission;
and, even when in emission, they are generally much weaker than 
the helium emission lines in the spectrum of CI~Cam \citep{Jasc98}.
The supergiant B[e] star Hen~S~134 does have strong He~I emission 
lines and its spectrum is strikingly similar to that of CI~Cam. 
[Compare the quiescent spectra of CI~Cam in \citet{Down84} 
and \citet{Orla00} to 
the spectrum of Hen~S~134 in \citet{Zick86}.]
Hen~S~134 is extremely luminous.
Even though it is in the LMC, its visual magnitude
is $V \approx 12.0$.
The de-extincted visual magnitude of CI~Cam is $V \approx 9.3$, so
if CI~Cam is as luminous as Hen~S~134, its distance is
greater than 10~kpc.

Also, as we will show in the next section, the Fe~II emission
lines in the spectrum of CI~Cam, which arise in a circumstellar wind
from the sgB[e] star,
have a half width at half maximum (HWHM) of only 32~km~s$^{-1}$.
The HWHMs of the Fe~II emission lines in most other sgB[e] stars are
typically 50 -- 75~km~s$^{-1}$ or more \citep{Oudm98}.
The notable exceptions are, once again, the extremely luminous
supergiant B[e] stars:
The Fe~II lines have a HWHM of 36~km~s$^{-1}$ in
Hen~S~134 and 20~km~s$^{-1}$ in the sgB[e] star
R126 (Zickgraf et al.\ 1985, 1986).
Since wind velocities tend to be strongly correlated with 
the surface gravities of the emitting stars, the
low velocity of the wind from CI~Cam again places it among the
largest and most luminous sgB[e] stars.

The second line of evidence is based on the structure of
the Galaxy.
The radial velocity of CI~Cam with respect to the Local
Standard of Rest is $-51$~km~s$^{-1}$ and its galactic longitude
and latitude are $(l,\; b) \approx (149^\circ,\; 4^\circ)$.
Since CI~Cam is a young, extreme Population~I object,
it should lie close to the plane of the galaxy and have a nearly 
circular orbit around the galactic center.
If we assume that its radial velocity is produced by
differential galactic rotation and use typical models for the 
galactic rotation \citep{Burt88a},
its distance is $\sim 7$~kpc.
Also, for any distances between 2 and 6~kpc, the galactic latitude
of CI~Cam places it several hundred parsecs above the plane of
the galaxy, much higher than expected for young objects.
However, the outer disk of the galaxy begins to warp upwards
in the direction of CI Cam at distances greater than a few
kiloparsecs, and at distances beyond 7~kpc the line of
sight to CI~Cam again penetrates the galactic disk
\citep{Burt88b}.
Thus the galactic latitude of CI~Cam implies distances 
approaching 7~kpc.

Finally, the known H~II regions with angular separations 
from CI~Cam less than $5^\circ$
have distances between $\sim0.9$ and $\sim9.0$~kpc
\citep{Blit82, Chan95}.
CI~Cam could reasonably lie at any distance up to $\sim9.0$~kpc.

For the purposes of this paper, it is sufficient to place CI~Cam at 5~kpc,
2.5 times further than previously supposed, but we emphasize that
this is likely to be a lower limit to the true distance.

\subsection{The X-ray Luminosity of CI~Cam}

According to \citet{Bell99} the unabsorbed
2 -- 25 keV flux from CI~Cam at the peak of its outburst was
$\sim 1.1 \times 10^{-8}\ \rm{erg\ cm}^{-2}\ \rm{s}^{-1}$.
At the revised distance, its X-ray luminosity becomes
\begin{equation}
   L(2-25\ \rm{keV})\ \approx \ 3.0 \times 10^{38} 
                          \left( 
                             {d \over {5\ \rm{kpc}}}
                          \right)^2\ 
                          \rm{erg\ s}^{-1}
\end{equation}
This luminosity lies near the upper end of the range of peak
luminosities for X-ray transients and is greater than the
luminosity of most of the black hole X-ray transients \citep{Chen97}.
CI~Cam was, therefore, a high-luminosity X-ray transient.

Using the quiescent hard X-ray flux measured by \citet{Orla00},
we find the quiescent hard X-ray luminosity of CI~Cam to be
\begin{equation}
   L(2-10\ \rm{keV})\ = \ 5.0 \times 10^{32} 
                          \left( 
                             {d \over {5\ \rm{kpc}}}
                          \right)^2\ 
                          \rm{erg\ s}^{-1}.
\end{equation}
Fits to the X-ray spectrum of CI~Cam in quiescence required a
second, extremely-soft, thermal bremsstrahlung component 
(kT $= 0.22\pm0.8$ keV) with an unabsorbed luminosity of
$2 \times 10^{34}\ \textrm{ergs s}^{-1}(d/5\ \textrm{kpc})^2$ in the
0.5 -- 2.0~keV band \citep{Orla00}.
While X-ray transients often display a soft X-ray excess during
outbursts, the excess is never this soft and never present
during quiescence.
We suspect this soft component is produced by the wind from
the sgB[e] star and is not directly associated with 
the compact object in CI~Cam.
Most high-luminosity OB stars are intrinsic X-ray sources and
their X-ray luminosity is strongly correlated with their bolometric 
luminosity and with the rate at which kinetic energy is carried away 
by their stellar winds.
According to \citet{Scio90}, their X-ray luminosity 
in the 0.2 - 4.0~keV band is
related to their bolometric luminosity by
\begin{equation}
  \log L_x \ = \ 1.08^{+0.06}_{-0.22} \log L_{bol} - 9.38^{+2.32}_{-0.83}.
\end{equation}
The bolometric luminosities of the sgB[e] stars range from
$\log (L_{bol}/L_\odot) = 5.0$ to 6.0 \citep{Zick98}.
For $\log (L_{bol}/L_\odot) = 5.5$
this relation allows X-ray luminosities approaching
$10^{34}\ \textrm{ergs s}^{-1}$,
so most of the quiescent X-ray flux from CI~Cam could be 
coming from the sgB[e] star.
\citet{Orla00} concluded that the quiescent X-ray flux 
cannot come from the sgB[e] star.
Our conclusion differs from theirs because we have adopted a
higher bolometric luminosity for CI~Cam and have used a more modern relation
between $L_x$ and $L_{bol}$.

Some of the flux from this extremely soft component in CI~Cam
may contribute to the flux in the 2.0 - 10~keV band and,
therefore, the quiescent flux observed in the 2.0 - 10~keV band is
an upper limit to the flux in that band from the compact star alone.
The ratio of quiescent luminosity to peak luminosity for the entire
CI~Cam system is 
$L_q / L_p \sim 1.7 \times 10^{-6}$, independent of distance,
but for the compact star alone, the ratio is
$L_q / L_p < 1.7 \times 10^{-6}$.

\section{The Kinematic Properties of the Circumstellar Environment}

The great range of the widths and profiles of the emission lines
in the spectrum of CI~Cam, and the
range of ionization states of the emitting species implies that 
the circumstellar
environment of CI~Cam is kinematically and thermally complex.
There are three kinematically distinct regions.
The first region is the source of the narrow, symmetric lines of 
[N~II] and [O~III], the second is the source of the metal lines 
and the hydrogen Paschen lines, and the third is the source of
the extremely broad lines of H, He~I, and Na~D.
The properties of the regions are summarized in Table~2.

\paragraph{Region I:}
The first region is characterized by emission lines with a
HWHM near 17~km~s$^{-1}$.
It is the source of the
highly-forbidden emission lines of [O~III] at 4363~\AA, 4959~\AA, and 
5007~\AA; and of [N~II] at  5755~\AA\ and 6583~\AA\ 
(the 6548~\AA\ line of [N~II] is hidden in the 
strong, steeply-sloped blue wing of H$\alpha$).
Figure~6 shows the [O~III] line at 5007~\AA\ and the [N~II] 
line at 5755~\AA.
The [O~III] lines have a nearly Gaussian profile with 
a HWHM of 16.4~km~s$^{-1}$.
The [N~II] line at 5755~\AA\ has two components, a
narrow component with a
Gaussian profile and a HWHM of 17.2~km~s$^{-1}$
and a broad component with a HWHM of 190~km~s$^{-1}$.
The broad component is displaced 15~km~s$^{-1}$
to the blue of the narrow component, causing 
the asymmetric line wings.
Only the narrow component is clearly visible in the 
[N~II] 6583~\AA\ line.
The radial velocities of the [O~III] lines and the
narrow component of the [N~II] lines are the same to within
the measurement error and their mean is
\( -43.2\ {\rm km\ s^{-1}} \).
No other lines in the optical spectrum, forbidden or otherwise, 
have similar kinematic properties.

\paragraph{Region II:}
The second region is characterized by emission lines with
HWHM between 32 and 35~km~s$^{-1}$.
This region is the source of all the permitted lines 
except for the Balmer lines, the helium lines, and the Na~D lines.
It is also the source of the [Fe~II] and [O~I] forbidden
lines and of most of the emission in the higher
members of the Paschen series.
The profiles of the weaker metal lines are nearly rectangular.
The stronger lines have more rounded profiles, becoming almost
Gaussian in shape as the lines become stronger, and
the very strongest lines have an extended blue wing.
The [Fe~II] lines are nearly rectangular but have a dip in the 
middle of their profiles.

We fit the emission lines in a portion of the spectrum between 
5415~\AA\ and 5436~\AA\ with a model in which the lines 
are the convolution of a rectangular line profile that has the 
same width for all lines, and a broadening profile that
can be different for different lines.
The physical picture behind this model is a spherically-symmetric, 
uniformly-expanding wind that is optically thin in the continuum
and produces emission lines.
The bulk motion of the wind gives an identical underlying rectangular 
profile to all the lines.
In addition, each line has its own local line profile.
The local line profiles all include a narrow Gaussian component, 
which is slightly wider for stronger lines than weaker lines;
and the profiles of the stronger lines have 
a second, broad Gaussian component to model their 
broad, asymmetric wings.
The convolution corresponds to a simple sum of the local
line profiles over the volume of the wind.

The fit of the model to the data, shown in Figure~7, is
excellent considering the model's simplicity.
The half width of the rectangular line profile is 0.58~\AA.
The 1-$\sigma$ width of the narrow Gaussian component
is 0.072~\AA\ for the four weakest lines and 0.117~\AA\ for the 
two strongest lines.
Broad Gaussian wings were needed for the three Fe~II lines 
but were not needed for the Fe~I, Ti~II, and Cr~II lines.
The width and displacement of the broad Gaussian component increased
with increasing line strength.

The expansion velocity of the wind, which we will call the ``iron wind,''
is given by the half width of the rectangular profile, 32~km~s$^{-1}$.
After a small correction for broadening in the spectrograph,
the width of the narrow Gaussian component 
in the four weakest lines is 3.1~km~s$^{-1}$, and, notably, its
width is the same in the Ti~II 5418~\AA\ and the Cr~II 5421~\AA\ lines 
even though they differ in strength by a factor of 2.2.
Since the widths and profiles of the weaker lines are independent 
of their strength, the profiles are dominated by kinematics, not 
radiative transfer.
In contrast, the profiles of the stronger Fe~II lines do depend
on line strength, so their profiles are affected by radiative transfer.

The iron wind is remarkably uniform.
The rectangular profiles of the weaker lines, which are
the best tracers of the wind geometry and kinematics, allow little
deviation from a spherically symmetric geometry and,
in particular, there cannot have been any large gaps in the wind.
There is no evidence that the iron wind rotates.
The projected rotational broadening of the lines must be a small
fraction ($\lesssim$10\%) of the expansion velocity to avoid
producing the double peaks and broad line wings that are typical
markers of rotation.
We will show in the next section that the width of the narrow Gaussian
component is due primarily to thermal broadening.
Thus, there is little turbulence and little radial acceleration
or deceleration in the wind.

\paragraph{Region III:}
The third region is characterized by emission lines with a
HWHM greater than 50~km~s$^{-1}$ and is 
the source of the Balmer lines, the He~I  lines, and the Na~D lines
(see Figures~1 and 8).
All these lines have broad, asymmetric wings and large equivalent
widths.
The He~I\ 6678\ \AA\ line and H$\beta$ are relatively
uncontaminated by other lines; we modeled them
with two Gaussian components: The line cores required Gaussians with a HWHM 
of $\sim 60\ {\rm km\ s^{-1}}$ for H$\beta$ and
$\sim 50\ {\rm km\ s^{-1}}$ for He~I; and
the broad wings of both lines required Gaussians with a HWHM of
$\sim 160\ {\rm km\ s^{-1}}$ displaced by 
$\sim 60\ {\rm km\ s^{-1}}$ to the blue.
Although a two-Gaussian fit does seem to work for all the lines
arising in this region, the widths of the Gaussians differ
from line to line.
For example, the HWHM of the lines in the Balmer series decrease from
$\sim 85\ {\rm km\ s^{-1}}$ at H$\alpha$ to
$\sim 50\ {\rm km\ s^{-1}}$ at H$\gamma$.

The full extent of the broad wings is difficult to measure because
the wings are overlaid by the thicket of weak emission lines,
but H$\alpha$ is so broad that it extends over more than one echelle
order and its blue wing extends to at least 6510\ \AA\, or 
$-2500\ {\rm km\ s^{-1}}$.
The wings of H$\beta$ extend from at least
4848 to 4872\ \AA\ or $-$800 to $+670\ {\rm km\ s^{-1}}$.
The strong variation  of the line widths and profiles 
with order in the Balmer series suggests that the lines
are strongly affected by radiative transfer.

The broad Balmer and the helium lines most likely come from 
high velocity outflow but our data place only weak constraints
on the the geometry of the outflow.
None of the emission lines in the visual and near-infrared region
have P-Cygni profiles, suggesting that the outflow is 
non-spherical and directed away from the line of sight.
On the other hand, the C~IV 1549~\AA\ and Si~IV 1394,1403~\AA\ 
lines seen our {\it HST} spectrum of CI~Cam do have 
pronounced P-Cygni profiles with absorption wings extending
to $\sim -1000\ \textrm{km\ s}^{-1}$, suggesting a more spherical outflow.
\citet{Iked00} obtained spectropolarimetry of CI~Cam
on 1998 April 4, 10, and 11, only a few days before we obtained 
our optical data.
The polarization at that time was entirely consistent with
interstellar polarization.
There was no evidence for intrinsic continuum polarization 
nor did the polarization change with wavelength over the profiles
of the strong emission lines,
again suggesting a more spherical geometry.
The data, then, slightly favor a weakly collimated
outflow -- a high velocity wind -- rather than a purely spherical
outflow or a narrowly collimated jet.
The velocity of the outflow is likely to be somewhat larger than
$2500\ {\rm km\ s^{-1}}$, the maximum observed velocity.

\section{The Thermal Properties of the Circumstellar Environment}

There are four thermally distinguishable regions around CI~Cam. 
Each of the three kinematic regions identified in the previous
section has distinct thermal properties.
The fourth region is the source of the infrared dust emission.

\paragraph{Region I:}
This region is the source of the highly forbidden [N~II]
and [O~III] lines.
Since the spectrograms of CI~Cam are not flux calibrated, we were forced
to measure the fluxes in these lines (and all other lines)
indirectly, by
first measuring their equivalent widths and then using
Johnson BVR magnitudes to calibrate the continuum.
All flux ratios were corrected for reddening.
The best estimate of the reddening is the one we have derived
from the ultraviolet extinction, $E(B-V) = 0.85 \pm 0.05$,
but we have also calculated line flux ratios for $E(B-V) = 0.65$ 
and $E(B-V) = 1.1$ to include possible effects of non-standard 
optical reddening.
There were many complications measuring the [N~II] lines.
The 6583~\AA\ line is difficult to measure because it lies in the 
strong and steeply-sloped red wing of H$\alpha$;
the 6548~\AA\ line is lost in the blue wing of H$\alpha$, forcing
us to deduce its flux from the flux in the 6583~\AA\ line and the 
ratio of the transition probabilities of the two lines; 
and the flux in the narrow component of the
5755~\AA\ line had to be deconvolved from the flux in the broad component.
Because of all these complications, the measured diagnostic line ratios 
have large uncertainties:
\begin{equation}
  \phantom{I} \textrm{[N~II]:} \phantom{mmmmm}
        0.47 \leq {{I(6548) + I(6583)} \over {I(5755)}}\ \leq\ 0.76
\end{equation}
\begin{equation}
 \textrm{[O~III]:} \phantom{mmmmm}
        5.4  \leq {{I(4959) + I(5007)} \over {I(4363)}}\ \leq\ 7.1 \\
\end{equation}
where the ranges include the range of reddening corrections.

We used the ``Nebular'' software package \citep{Shaw95} to determine the
temperature and density of the gas producing the [N~II] and [O~III]
lines.
The allowed values of $\log n_e$ and $\log T_e$ are shown in Figure 9.
Both sets of lines must come from regions with high electron densities,
$\log n_e > 5.8$, but the temperatures could be fairly low, $\log T_e > 3.85$.
One does not normally expect the [N~II] and [O~III] lines to come
from the same volumes of gas because their ionization potentials
are so different, but the similarity of their line profiles
suggests some overlap of their line formation regions.
If so, the temperature and density of the overlap region 
is given by the overlap area in Figure~9:
\begin{eqnarray}
   \log n_e & = & 6.2 \pm 0.3 \nonumber \\
   \log T_e & = & 4.3 \pm 0.2
\end{eqnarray}
These temperatures and densities are high enough that physical effects
not included in the Nebular package could be affecting the line ratios,
e.g., radiative and dielectronic recombinations,
but the uncertainties introduced by these other effects should be smaller
than the large uncertainties introduced by the measurement errors.

\paragraph{Region II:}
The iron wind is the second thermal region.
The wind emits both Fe~II and [Fe~II] lines, but the
forbidden lines are generally quite weak
(the 5158 \AA\ line shown in Figure~2 is
a rare example of a fairly strong [Fe~II] line).
The density in the wind is, then, comparable to or
greater than the critical density
for depopulating the metastable states by collisions.
According to \citet{Netz88} and \citet{Vern99} the critical
density is $\log n_e \approx 9.5$.
Because of this high density, we can also place a meaningful upper
limit on the electron temperature in the wind without a
detailed treatment of heating and cooling processes.
The wind does not produce any Fe~III emission lines and, therefore,
electron collisions do not
ionize a significant fraction of the Fe~II atoms to Fe~III.
From the Saha equation,
the kinetic temperature of the boundary between the two
stages of ionization is roughly given by $kT_e \sim E_{3,2} / \gamma$
where $E_{3,2}$ is the ionization potential of Fe~II and
$\gamma = 35.4 - \ln n_e + 1.5 \ln T_e$ \citep{Rybi79}.
Taking $\log n_e = 10$, we find $\gamma \approx 23.5$ and the
upper limit to the temperature is $T_e < 8000~K$.
The presence of Fe~I lines from the wind (e.g., Multiplet 15)
supports this rather low temperature.
Conversely, the temperature of the iron wind must be
greater than $\sim 7000~K$ to produce Fe~II emission lines \citep{Netz88}.
Thus, the temperature of the wind is close to $8000~K$.

In the previous section we found that the local Doppler broadening
of the emission lines coming from the wind is 3.1~km~s$^{-1}$.
The thermal velocity of iron atoms at $8000~K$ is about 1.9~km~s$^{-1}$,
so all other sources of Doppler broadening, including turbulent velocities 
and gradients in the expansion velocity, amount to only 2.4~km~s$^{-1}$.
The sound speed in gas with a temperature of $8000~K$ is $\sim 9$~km~s$^{-1}$
so any turbulent flow of gas within the wind is certainly subsonic.
There is remarkably little velocity structure to the wind.

We can roughly estimate the mass loss rate $\dot M$ in the iron
wind from
\begin{equation}
 \dot M \ = \ 4 \pi r_w^2 \rho v
        \ \approx \ \left( {{L_*} \over {4 \sigma T_w^4}} \right)
              \left( {{n_e \mu m_H} \over { f }} \right)
              v,
\end{equation}
where $r_w$ is a characteristic radius from which most of the
Fe~II emission is coming, 
$\rho$ is the mass density of the wind, $v$ is the velocity of the
wind, $L_*$ is the luminosity of the sgB[e] star, $T_w$ is the
temperature of the wind, $f$ is the fraction of hydrogen
that is ionized, and the remaining symbols have their usual meanings.
We take $r_w$ to be radius where
the energy density of the diluted stellar radiation field has an
equivalent black body temperature equal to the wind temperature,
$4 \pi r_w^2 \sigma T^4_w = L_*/4$.
Taking $L_* = 10^5\ L_\odot$, $T_w = 8000~K$, $\log n_e = 10$,
and $v = 32$~km~s$^{-1}$, we find
$\dot M = 4.4 \times 10^{-7}\; (1 /f)\ M_\odot\ \textrm{yr}^{-1}$.

\paragraph{Region III:}

The third thermal region is the source of the strong emission lines
of hydrogen and neutral helium.

Most of the helium lines are not suitable for a simple analysis
because they are heavily blended and difficult to measure accurately,
and they are likely to be optically thick and subject to the effects of
radiative transfer.
Exceptions are the weak lines at 4713~\AA\ and 5048~\AA.
Both are 2p-4s transitions, the 5048~\AA\ line in
the singlets and the 4713~\AA\ line in the triplets, and
should be much less optically thick than the stronger helium lines;
and both lines are also relatively uncontaminated by blends.
The ratio of their de-reddened intensities is
\begin{equation}
  { {I(4713)} \over {I(5048) }}\ =\  3.8 \pm 0.4
\end{equation}
Unfortunately, the very weakness of these lines means the measured
line intensities may have substantial internal and
external errors, precluding a precise quantitative analysis.
Nevertheless, the
recombination model and computer program ``he5'' \citep{Benj99}
suggests that the gas emitting the lines has an electron temperature
near $15,000\ K$ with a wide range of possible electron densities centered
near $\log n_e = 6$.

The hydrogen Balmer lines have a large decrement.
The ratios of their dereddened fluxes at the peaks of the lines are
\begin{equation}
   {{\textrm{H}\alpha} \over {\textrm{H}\beta}} \ = \ 5.3\pm1.3
   \ \ \ \ \textrm{and} \ \ \ \ 
   {{\textrm{H}\gamma} \over {\textrm{H}\beta}} \ = \ 0.35 \pm 0.03,
\end{equation}
which is much steeper than the decrement for Case~B recombination
$(H_\alpha/H_\beta/H\gamma \approx 2.8/1.0/0.47)$.
There are two basic ways to produce steep Balmer decrements \citep{Drak80}.
The first is to have a large electron density,
$\log n_e \approx 10 - 12$, so that electrons are excited into
the upper levels of the hydrogen
atoms by collisions.
This mechanism is inconsistent with the densities implied
by the helium line ratios.
The second is to have a low electron density, $\log n_e < 7$,
but a high optical depth to H$\alpha$, $\tau_{H\alpha} > 100$,
so that Balmer photons are trapped and rapidly degrade to
lower members of the Balmer series.
This mechanism is consistent with the helium line ratios.
We conclude that the steep decrement is produced by
high optical depth in H$\alpha$.

\paragraph{Region IV:}
The infrared emission from CI~Cam comes from a fourth 
thermally-distinct region.
The infrared emission, which was present before, during, and after
the 1998 outburst \citep{Alle73, Bell99, Clar00}, 
is produced by dust in the circumstellar envelope.
The spectral energy distribution of the dust emission has been discussed
at length by \citet{Bell99} and \citet{Clar00};
here we discuss only the size of the dust shell.

We assume that the infrared emission comes from dust that
condenses from the stellar wind when the wind cools below
the condensation temperature.
Most of the emission comes from the hottest dust
near the inner edge of the dusty shell.
We assume that this dust has a temperature near $T_{\rm d} = 1350~K$
and that the inner edge is located at a 
characteristic distance $r_{\rm shell}$ from CI~Cam.
If the dust is heated by the sgB[e] star in CI~Cam and is in 
thermal equilibrium, then
\begin{equation}
c \int {\rm C_a}(\lambda) \; U_\lambda \; d\lambda\ =\ 
          4\pi  \int {\rm C_e}(\lambda) \; B_\lambda(T_{\rm d})\; d\lambda
       \label{dust1}
\end{equation}
where 
$c$ is the speed of light,
$U_\lambda$ is the radiation energy density, and
${\rm C_a}(\lambda)$ and ${\rm C_e}(\lambda)$ are the 
absorption and emission
cross sections of the dust grains [e.g., \citet{Doty94}].
We take $c U_\lambda = 4 \pi W B_\lambda(T_*)$, where 
$T_*$ is the effective temperature of sgB[e] star,
$W$ is the geometric dilution factor,
\begin{equation}
     W \ = \ {{ R^2_*} \over {4r^2_{shell}}},
    \label{dust2}
\end{equation}
and $R_*$ is the effective radius of the sgB[e] star.
We have, then,
\begin{equation}
 W \int {\rm C_a}(\lambda) \; B_\lambda(T_*) \; d\lambda\ =\ 
            \int {\rm C_e}(\lambda) \; B_\lambda(T_{\rm d})\; d\lambda
       \label{dust3}
\end{equation}
We take ${\rm C_a}\ =\ {\rm C_e}\ =\ {{\rm C_o}/ \lambda^p}$,
where $p = 0$ corresponds to efficient radiative cooling of the
grains and gives a lower limit to $r_{shell}$;
and $p = 1$ corresponds to inefficient cooling and gives an
upper limit to the radius.
For $p = 0$, equations~\ref{dust2} and \ref{dust3} yield
\begin{equation}
  r^2_{shell} \ = \ {{L_*} \over {16 \pi \sigma T^4_{\rm d} } }
  \label{dust4}
\end{equation}
and for $p = 1$, they yield
\begin{equation}
  r^2_{shell}\ =\ { {L_*T_*} \over {16 \pi \sigma T^5_{\rm d} } }.
    \label{dust5}
\end{equation}
where $L_*$ is the luminosity of the sgB[e] star.
For $L_* = 10^5 L_\odot$ and $T_* = 20,000~K$, equations~\ref{dust4}
and \ref{dust5} give
\[
    13~{\rm AU} \ < \ r_{shell} \ < \ 52~{\rm AU}.
\]
If the distance to CI~Cam is 5~kpc, the angular radius of the shell
is between 0.0026 and 0.0104 arcseconds, in agreement
with the results of \citet{Trau99}, who found an angular diameter
of 0.0052 arcsec in the K band using the IOTA interferometer.

Using the scaling relation from \citet{Bjor98}, a spherically
symmetric wind can produce dust particles only if the rate of mass loss
in the wind is greater than
\[
   \dot M\ > \ 1.7 \times 10^{-7} 
             \left({r_{shell} \over {\textrm{25 AU}}} \right)
             \left({v_\infty \over {30\ \textrm{km s}^{-1}}} \right)^2
             M_\odot\ \textrm{yr}^{-1}
\]
where $v_\infty$ is the wind speed.
Adopting $v_\infty = 32\ \textrm{km s}^{-1}$ and
$r_{shell} = 25$~AU, 
we find $\dot M > 1.9 \times 10^{-6}M_\odot\ \textrm{yr}^{-1}$
in good agreement with the mass loss rate deduced for the
Fe~II wind if the ionization fraction $f$ is less than
$\sim 0.1$.
Unlike \citet{Clar00}, we conclude that the wind 
need not be episodic nor confined to the equatorial plane of the 
sgB[e] star.

\section{Discussion}

\paragraph{The Environment of the Compact Star in CI~Cam.}

The compact star in CI~Cam is immersed in a complex environment 
produced by a two-component wind from the sgB[e] star.
One component is a cool, low-velocity wind (the ``iron wind''),
and the other is a hot, high-velocity wind.
The iron wind is dense, $\log n_e > 9.5$, roughly spherical,
continuously replenished, and carries away mass at a high
rate, $\dot M > 10^{-6}\ M_\odot\ \textrm{yr}^{-1}$.
The wind fills the space around the sgB[e] star and, based
on the radius of the inner edge of the infrared-emitting dust shell, 
it extends to at least 13~AU.
The hot, high-velocity wind is the source of the broad H, He, and 
Na emission lines in the spectrum of CI~Cam.
This wind has a velocity in excess of 2500~km~s$^{-1}$, a temperature
of $1.7\pm0.3 \times 10^4~K$, and a density
of $\log n_e \approx 6$.

It is far from clear how the high-velocity and low-velocity winds
can co-exist in CI~Cam.
Other sgB[e] stars also show two-component winds.
Typical models 
for their winds make the low-velocity wind an equatorial outflow, 
allowing the high velocity wind to escape as a polar outflow.
Such models cannot be entirely correct for CI~Cam because
the low-velocity wind is spherical, not confined to a plane.
The high-velocity wind was, however, enhanced and changing rapidly 
when we observed it.
The broad Na D lines present during the outburst were not 
present before or after the outburst, and although H and He lines 
are always present, they became much stronger during the outburst
\citep{Bars98, Orla00, Jasc00}.
It is possible that the
enhanced high-velocity wind may eventually interact with the
material in the low velocity wind
but had not done so at the time of our observations.

As noted by \citet{Bell99} and others, there are
similarities between CI~Cam and the Be-star X-ray transients.
A typical Be-star X-ray transient contains a neutron star in an
eccentric orbit that carries it once per orbit through a 
wind confined to the equatorial plane around the Be star.
Enhanced accretion during passage through the equatorial wind causes
periodic X-ray outbursts \citep{Bild97}.
The outbursts are variable in duration and luminosity.
In A0538-66, which has an orbital period of 16.65 days,
the outbursts can last from a few hours to 10 days and can
have luminosities between $10^{37}$ and $10^{39}\ \textrm{erg\ s}^{-1}$
\citep{Corb97}.
The resemblance between CI~Cam and the Be-star X-ray transients
is only superficial, however.
CI~Cam is a sgB[e] star, emphatically not a Be star, and its
circumstellar material is much denser, far more extended,
and much less confined to the equatorial plane than the
circumstellar material around a Be star.
It is more correct to think of the compact star in CI~Cam as
continuously burrowing through dense circumstellar material 
as it orbits the sgB[e] star.
The dense environment of the compact star makes CI~Cam 
unique among the known X-ray binaries.

\paragraph{The Peculiarly Low X-ray Luminosity at Quiescence.}

Because the compact star continuously burrows through the
dense wind from the sgB[e] star,
accretion is not naturally limited to a short
interval of orbital phases as in the Be-star/neutron-star systems.
The reverse is true:
The compact star should always be accreting at a high rate, 
even during quiescence.
Let us assume for the moment that the compact star accretes
material from the wind at the Bondi accretion rate.
For a compact star of mass $M_x$ moving with relative velocity $V$
through a gas with an undisturbed density $\rho_\infty$ and
sound speed $c_s$, the Bondi accretion rate is
\citep{Bond52}:
\begin{equation}
  \dot M_x\ \approx\ 2 \pi\rho_\infty\; (GM_x)^2
        \left( V^2 + c_s^2 \right)^{-3/2}.
\end{equation}
The electron number density in the iron wind is $\log n_e > 9$, so even
if the wind is fully ionized, the mass density is greater than
$\rho_\infty = 3.3 \times 10^{-15}\ \textrm{gm\ cm}^{-3}$.
If the sum of the masses of the two stars is 
$M_B + M_x = 35 M_\odot$ and the semi-major axis of the
relative orbit is $a = 0.5$~AU, 
the orbital period is $2.0 \times 10^6$~s and the
velocity of the compact star in its relative orbit
around the sgB[e] star is $230\ \textrm{km~s}^{-1}$.
The iron wind around CI~Cam is expanding at $32\ \textrm{km s}^{-1}$.
Adding the wind velocity and the orbital velocity in quadrature,
we find $V \approx 230\ \textrm{km s}^{-1} >> c_s$.
If the compact star is a neutron star with a mass $M_x = 1.4M_\odot$, the
Bondi accretion rate is then
$\dot M_x = 6 \times 10^{16}\ \textrm{gm s}^{-1}$
and the Bondi accretion luminosity is
$\dot E \approx 7 \times 10^{36}\ \textrm{erg s}^{-1}$.
If the compact star is a black hole with a mass $M_x = 5 M_\odot$,
the accretion rate is 
$\dot M_x = 8 \times 10^{17}\ \textrm{gm s}^{-1}$
and the accretion luminosity is
$\dot E \approx 4 \times 10^{37}\ \textrm{erg s}^{-1}$.
The luminosity scales as $a^{3/2}$.

The Bondi accretion luminosity
differs greatly from the observed quiescent X-ray luminosity,
$L(2 - 10\ \textrm{keV}) = 5 \times 10^{32}\; 
(\textrm{d}/5\ \textrm{kpc})^2\ \textrm{erg\ s}^{-1}$.
The fundamental reason for the high luminosity is the large 
capture radius for Bondi accretion:
\begin{equation}
  r_{cap} \ \approx \ {{2 G M_x} \over {V^2}}.
\end{equation}
For the example at hand,
$r_{cap} = 0.05$~AU for a $1.4M_\odot$ neutron star and
$r_{cap} = 0.17$~AU for a $5M\odot$ black hole.
Some process either prevents the compact star from accreting at
the Bondi rate or prevents the accretion energy from being
emitted at X-ray energies.
Whatever the process, it cannot rely on a fine tuning
of parameters to prevent the X-ray emission because the material
in the circumstellar envelope has a wide range of temperatures, 
densities, and velocities.

\paragraph{The Nature of the Compact Star and the Source of the Outburst.}

The most direct ways a neutron star can betray itself is through
Type~I bursts, which are caused by a thermonuclear runaway in
the envelope of a neutron star, or
through type~II bursts or periodic modulations of the X-ray light
curve, which require a magnetized neutron star.
None of these has been observed in CI~Cam.
Even quasi-periodic oscillations are absent.
\citet{Bell99} have suggested that periodicities
from a neutron star could be smeared or obliterated by
strong absorption or scattering of the X-rays,
but the X-ray spectrum is inconsistent with strong absorption
or scattering.
Thus, there is no direct evidence for a neutron star in CI~Cam.

Although there is likewise no direct evidence for a black 
hole, the high X-ray luminosity of CI~Cam during outburst,
$L(2-25\ \rm{keV}) > 3.0 \times 10^{38}\ \rm{erg\ s}^{-1}$,
and the large ratio of its quiescent to peak luminosities,
$L_q/L_p < 1.7 \times 10^{-6}$, do provide indirect evidence
for a black hole.
Figure~10 is a modified version of similar figures in \citet{Nara97} 
and \citet{Garc98}.
The figure plots $L_q/L_p$ against $L_p$ for the X-ray
novae in which the compact star has been positively identified
as a neutron star or black hole.
The systems with neutron stars are shown as open circles and those
with black holes are shown as filled circles.
With only one exception the black holes systems have higher peak
luminosities than the neutron star systems, and with only one 
possible exception
the ratio of quiescent to peak luminosities of the black hole systems
is smaller than the that of the neutron stars systems.
With \emph{no} exceptions the neutron star systems are confined to
the upper left quadrant of the figure and are fully segregated from
the black hole systems.

\citet{Nara97} and \citet{Garc98} originally
proposed two reasons for the segregation.
First, black holes have higher peak luminosities because they
are more massive than neutron stars and have higher Eddington
luminosities.
Second, any residual accretion onto a black hole during quiescence
is swallowed by the event horizon and produces little X-ray emission,
whereas residual accretion onto a neutron star hits the surface
of the neutron star and does produce emission.
The real reasons for the segregation are likely to be more complicated
\citep{Meno99} but the empirical fact remains that neutron stars 
and black holes are segregated in the figure.

CI~Cam is also plotted in Figure~10.
We have adopted a peak luminosity corresponding to a distance 
of 5~kpc, the lower limit to the true luminosity, and a 
luminosity ratio corresponding to all the quiescent luminosity 
coming from the compact star, an upper limit to the
true ratio.
Even taking these limits,
CI~Cam falls within the region occupied by the black holes and
outside the region occupied by the neutron stars.
Figure~10 is, of course, circumstantial evidence for a black hole, not
proof, but the evidence is strong enough to make CI~Cam
a good candidate for a black hole binary.

We speculate that the X-ray outburst of CI~Cam was caused by the same
mechanism responsible for the outbursts of X-ray novae,
that is, by an instability in the accretion 
disk around the compact star \citep{Laso01}.
We further speculate that the reason the outburst was so short
is that the accretion disk in CI~Cam is smaller than the accretion
disks in other X-ray novae because the disk is fed from a
stellar wind, not by Roche-lobe overflow.
The rate of decline of a disk-instability outburst 
should scale by the viscous time scale in the disk,
\begin{equation}
  \tau \ = \ {r \over v_r} \ \propto \ r^{7/5}, \label{disk1}
\end{equation}
where $v_r$ is the velocity at which matter drifts radially 
inward due to viscosity \citep{Fran92}.
The exponent on $r$ corresponds to electron scattering
opacity, which is more appropriate than free-free opacity
for the outburst state.
The outburst of CI~Cam decayed on an e-folding time scale between
0.6 and 2.3 days; in contrast,
the Fast Rise Exponential Decay (FRED) outbursts of X-ray novae 
generally have decay timescales between 10 and 50 days
\citep{Chen97}.
Equation~\ref{disk1} implies that the disk in CI~Cam is 5 to 20 
times smaller than the disks in more typical X-ray transients.

The radius of an accretion disk produced by Roche-lobe overflow
should should be about 1/4 of the separation of the stars.
We take A0620-00 and V404~Cyg as typical X-ray novae with
FRED outbursts and with
accretion disks maintained by Roche-lobe overflow.
Their orbital periods are 0.323 d and 6.46 d respectively, 
and the radii of their accretion disks should be roughly 
$6 \times 10^{10}$~cm and $60 \times 10^{10}$~cm.
The disk in CI~Cam should be 5 to 20 times smaller than these
disks.

Wind accretion can produce the smaller disk in CI~Cam.
The radius of an accretion disk maintained by wind accretion is
given roughly by the circularization radius,
\begin{equation}
   r_{circ} \ = \ {{G^3M_x^3\; \omega^2} \over {V^8}},
\end{equation}
where $\omega \approx 2\pi / P_{orb}$
and $P_{orb}$ is the orbital period of the compact star
\citep{Fran92}.
For the example of the previous section $P_{orb} = 2.04 \times 10^6$~s
and $V = 230\ \textrm{km\ s}^{-1}$, yielding
$r_{circ} \approx 7.9 \times 10^{8}\ \textrm{cm}$ 
for a $1.4M_\odot$ neutron star, and
$r_{circ} \approx 3.6 \times 10^{10}\ \textrm{cm}$ 
for a $5M_\odot$ black hole.
Despite the large exponent on $V$, $r_{circ}$ scales only
linearly with the semi-major axis of the relative orbit.

Thus, unless the mass of the compact star is significantly greater
than $5M_\odot$ or the semi-major axis of its orbit is greater
than 0.5~AU, the size of the wind-fed accretion disk is small enough
to account for the short outburst.

Finally, we note that the black hole binary V4641~Sgr also had 
anomalously-short but high-luminosity X-ray outbursts.
The orbital period of V4641~Sgr is 2.8 days and the secondary
star in the system is an evolved B-type star with a mass
between 5.5 and $8.1 M_\odot$ \citep{Oros01}.
The black hole in this system could also be accreting via
a small disk that is fed by a wind or a focused wind from the B star
instead of a stream through the inner Lagrangian point.

\section{Summary}

We obtained high dispersion spectroscopy of CI~Cam, the optical
counterpart of XTE~J0421+560, two weeks after the peak of
its short outburst in 1998 April.  
The salient results from the observations are:

\begin{enumerate}

\item CI~Cam is a sgB[e] star. 
Its spectrum has a blue continuum with
strong, broad emission lines of H, He, and Na that were greatly
enhanced during the outburst.
It also has a thicket of weak, narrow emission lines of neutral and
singly ionized metals, and even narrower lines of [N~II] and [O~III].
There are no absorption lines in the visual spectrum other than
interstellar absorption features.

\item Based on the ultraviolet continuum between 1150~\AA\ and 2700~\AA, 
the reddening and extinction are
$E(B-V) = 0.85 \pm 0.05$, $A_V = 2.3\pm 0.3$, and $R_V = 2.7 \pm 0.2$.

\item CI~Cam is more distant than previously supposed.
Its spectroscopic properties place it among the
most luminous supergiant B[e] stars and imply a distance
greater than 5~kpc.
The radial velocity of CI~Cam with respect to the Local Standard of
Rest, $-51$~km~s$^{-1}$, and its galactic coordinates,
$(l,\; b) \approx (149^\circ,\; 4^\circ)$, also imply a distance
greater than 5~kpc.

\item The X-ray luminosity at the peak of the outburst was
$L(2-25\ \rm{keV})\ \approx \ 3.0 \times 10^{38}\ \rm{erg\ s}^{-1}$
for a distance of 5~kpc,
making CI~Cam one of the most luminous X-ray transients.
The ratio of quiescent luminosity to peak luminosity in the 2 -- 25~keV
band is $L_q/L_p < 1.7 \times 10^{-6}$.

\item The sgB[e] star emits a two-component wind.
One component is a cool, low-velocity wind (the ``iron wind''),
which is dense, $\log n_e > 9.5$, roughly spherical,
continuously replenished, and has been present since at least 1931.
The mass loss rate due to the wind is high, 
$\dot M > 10^{-6}\ M_\odot\ \textrm{yr}^{-1}$.
The wind fills the space around the sgB[e] star and, from
the size of the infrared-emitting dust shell, extends
to a radius between 13~AU and 50 AU.
The second component is a hot, high-velocity wind, which is
the source of the broad H, He, and Na emission lines in the 
spectrum of CI~Cam.
This wind has a velocity in excess of 2500~km~s$^{-1}$, a temperature
of $1.7\pm0.3 \times 10^4~K$, and an electron number density
of $\log n_e \approx 6$.
It is unclear how the high-velocity and low-velocity winds can
co-exist.

\item Although the compact star in CI~Cam is immersed in the dense
circumstellar wind from the sgB[e] star, it 
burrows through the wind while producing only a small fraction,
$< 10^{-4}$, of the X-ray emission expected from the Bondi accretion rate
except for rare transient outbursts.
This picture, a compact star traveling in a wide orbit
through the dense circumstellar envelope of a sgB[e] star,
occasionally producing transient X-ray outbursts, makes
CI~Cam unique among the known X-ray binaries.

\item The lack of type~I and type~II bursts, the lack of periodic
X-ray pulsations, the great luminosity at the peak of the outburst,
and the large ratio of peak to quiescent luminosity is circumstantial
evidence that the compact object in CI~Cam is a black hole, not
a neutron star.

\item We speculate that the outburst of CI~Cam was caused by the same
disk instability mechanism responsible for the outbursts of X-ray novae.
The reason the outburst was so short
is that the accretion disk in CI~Cam is smaller than the accretion
disks in other X-ray novae because the disk is fed from a
stellar wind, not by Roche-lobe overflow.

\end{enumerate}

\acknowledgments{We thank H. Dinerstein, R. Hynes, Z. Ioannou, A. King,
G. Shields for useful discussions.  Support for proposal \#8138 was
provided by NASA through a grant from the Space Telescope Science
Institute, which is operated by the Association of Universities for
Research in Astronomy, Inc., under NASA contract NAS 5-2655.
}

\begin{deluxetable}{clcl}

\tablecolumns{4}
\tablewidth{4.5in}
\tablecaption{Identified Lines in the Spectrum of CI Cam}
\tablehead{
  \multicolumn{2}{c}{Permitted Lines}  &  \multicolumn{2}{c}{Forbidden Lines} \\
  \cline{1-2} \cline{3-4}  \\
\colhead{Species} & \colhead{Example~~} & 
\colhead{Species} & \colhead{Example~~} }
\startdata
      H      &   Balmer series &    [N II]   &  5755 \AA        \\
      He~I   &   5876 \AA      &     [O I]   &  6300 \AA        \\
      He~II  &   4686 \AA      &    [O III]  &  4959/5007 \AA   \\
      C~I    &   Multiplet 3   &    [Fe II]  &  5158 \AA        \\
      N~I    &   Multiplet 1   && \\
      O~I    &   8446 \AA      && \\
      Na~II  &   D lines       && \\
      Mg~I   &   Multiplet 2   && \\
      Si~II  &   Multiplet 5   && \\
      Ca~II  &   IR triplet    && \\
      Sc~II  &   Multiplet 29  && \\
      Ti~II  &   Many          && \\
      Cr~II  &   Many          && \\
      Fe~I   &   Multiplet 15  && \\
      Fe~II  &   Many          && \\
\enddata
\end{deluxetable}

\clearpage

\begin{deluxetable}{llll} 

\tablecolumns{4}
\tablewidth{5.0in}
\tablecaption{The Kinematic Properties of the Circumstellar Environment}
\tablehead{
\colhead{} & \colhead{Region I} & \colhead{Region II} & \colhead{Region III} }
\startdata
 Typical         &[N II], [O III] & Metal lines        & Balmer lines   \\
  \quad lines    &                & Paschen lines      &  He I, Na~D      \\
  & & & \\
 Half Width at   & 16~km~s$^{-1}$ & 32~km~s$^{-1}$     & 50 -- 85~km~s$^{-1}$ \\
 \quad Half Max  &                &                    &                \\
  & & & \\
 Line            & Roughly~~~~    &  Weak lines        & Asymmetric;~  \\
 \quad Profiles  & \quad Gaussian &  \quad rectangular;& Blue wings to  \\
                 &                &  Strong lines      & ~$-$2500~km~s$^{-1}$\\
                 &                &  \quad asymmetric  & Red wings to   \\
                 &                &  \quad Gaussians   & ~$+$1000~km~s$^{-1}$\\
 \\
\enddata
\end{deluxetable}

\clearpage

\begin{deluxetable}{cccccc} 

\tablecolumns{6}
\tablewidth{6.5in}
\tablecaption{The Thermal Properties of the Circumstellar Environment}
\tablehead{
 \colhead{}  & \colhead{Region I} & \colhead{Region II} & \colhead{Region III} & \colhead{Region IV} }
\startdata
  Source of:            &  Highly   &  Metal  &   Balmer    & Infrared  \\
                        & forbidden &  lines  &  \& He I    & continuum \\
                        &  lines    &         &   lines     &           \\
                        &           &         &             &           \\
 Velocity (km~s$^{-1})$ &    16     &     32  &  Up to 2500 &     ...   \\
                        &           &         &             &           \\
 Temperature ($K$)      &    ...    &    8000 & 
                                     $(1.7\pm0.3) \times 10^4$ & 1350 K \\
                        &           &         &             &           \\
$\log n_e$ (cm$^{-3})$  &   $>$6.0  & $>9.5$  & $\approx 6$ &    ...    \\
 \\
\enddata
\end{deluxetable}

\clearpage

\begin{center}
  \textbf{Figure Captions}
\end{center}

\figcaption[nahe.ps]{
The spectrum of CI~Cam from 5825~\AA\ to 5910~\AA.
The He~I line at 5876~\AA\ has an equivalent width of 
$110\pm15$~\AA\ and broad asymmetric wings extending further 
to the blue than the red.
The Na~D lines at 5890/96~\AA\ are
cut by at least two complexes of interstellar absorption.
The spectrum is magnified in the lower panel to show 
the full extent of the blue wing on the He~I line, which extends to
at least 5845~\AA\ (--1500~km~s$^{-1}$), and the 
myriad of weak metal lines.
The weak absorption at 5850~\AA\ is a diffuse interstellar
absorption band.
\label{fig1}
}

\figcaption[idents.ps]{
The spectrum of CI~Cam from 5125~\AA\ to 5205~\AA\ with identifications
for the more prominent emission lines.
Although the strongest lines in this region of the spectrum come from
permitted Fe~II, lines of Ti~II, Mg~I are also present.
Lines of [Fe~II] are generally weak in the spectrum of CI~Cam;
the lines at 5158.0/58.8~\AA\ are rare examples of fairly strong lines.
\label{fig2}
}

\figcaption[ir1.ps]{
The spectrum of CI~Cam from 8380~\AA\ to 8510~\AA.
The strongest lines are the O~I line at 8446~\AA\ and
the Ca~II line at 8498~\AA, one of the calcium infrared triplet
lines.
The spectrum is magnified in the lower panel to show the
Paschen lines.
The Paschen lines can be identified to P34 in the next order of
the echelle spectrogram.
\label{fig3}
}

\figcaption[ir2.ps]{
The spectrum of CI~Cam from 8590~\AA\ to 8725~\AA.
The strongest line is the Ca~II line at 8662~\AA, a member
of the calcium infrared triplet.
The spectrum is magnified in the lower panel to show the
weaker lines in the spectrum, notably the N~I lines from
multiplets 1 and 8, the P14 line of hydrogen, 
and both permitted and forbidden lines of Fe~II.
\label{fig4}
}

\figcaption[redden.ps]{
The lower panel is the observed, heavily extincted, 
ultraviolet spectrum of CI~Cam.
The original spectrograms have been rebinned into 15~\AA\ bins
in the short wavelength region and 20~\AA\ bins in the long
wavelength region.
The upper pannel shows the de-extincted spectrum of CI~Cam 
with the spectrum of BD$+33^\circ2642$ overplotted for comparison.
The residual differences between the CI~Cam spectrum 
and the BD$+33^\circ2642$ 
spectrum are all due to real features in the CI~Cam spectrum:
Fe~II emission between 2300~\AA\ and 2500~\AA\ and wind
features at shorter wavelengths, for example.
\label{fig5}
}

\figcaption[onlines.ps]{
The highly forbidden lines of [O~III] at 5007~\AA\ and
[N~II] at 5754~\AA.
The half width at half maximum of both lines is
16~km~s$^{-1}$; the [N~II] line also has broad wings
extending to $\pm 5$~\AA\ ($\pm260\ \textrm{km~s}^{-1}$).
\label{fig6}
}

\figcaption[5426plot.ps]{
The solid line is the observed spectrum of CI~Cam from 
5416~\AA\ to 5436~\AA.
The lines of Ti~II, Cr~II, and Fe~I are symmetric and
rectangular with a pronounced flat top;
the Fe~II lines are more rounded and have asymmetric wings.
The half width at half maximum of all the lines is close
to 32~km~s$^{-1}$.
The dashed line is the spectrum produced by a model in
which each line is the convolution of a rectangular
profile and a line broadening profile -- a narrow Gaussian for
the lines of Ti~II, Cr~II, and Fe~I, plus a second
broader gaussian for the Fe~II lines to represent their asymmetric 
wings.
\label{fig7}
}

\figcaption[balmer.ps]{
The H$\alpha$ and H$\beta$ lines of hydrogen in CI~Cam.
The equivalent width of H$\alpha$ is $\sim 750\pm50$~\AA\ and that
of H$\beta$ is $73\pm3$~\AA.
The spectra are magnified in the lower panel to show the
wings of the lines.
The blue wing of H$\alpha$ extends to at least 6510~\AA\
(--2400~km~s$^{-1}$) and the blue wing of H$\beta$ to
4845~\AA\ (--1000~km~s$^{-1}$).
\label{fig8}
}

\figcaption[nete.ps]{
The locus of points in the $(\log n_e,\; \log T_e)$ plane
corresponding to the [O~III] (the dashed lines) and [N~II] 
(the solid lines) emission line ratios.
The upper line of each pair corresponds to a reddening
of $E(B-V) = 1.1$ and the lower line corresponds to
   $E(B-V) = 0.65$.
\label{fig9}
}

\figcaption[luminosity.ps]{
The outburst luminosities of X-ray novae and their luminosity ranges
between outburst and quiescence.
Neutron star systems are denoted by open circles, black hole systems
by small filled circles, and CI~Cam by the large filled hexagon.
The downward pointing arrows denote systems for which the
quiescent flux has not been measured.
CI~Cam is plotted at an outburst luminosity corresponding to a distance of
5~kpc, the lower limit to the true luminosity.
The neutron star systems are confined to the upper left quadrant of the
figure; CI~Cam is in the lower right quandrant where most of the black
hole systems are located.
}

\newpage
\epsfig{file=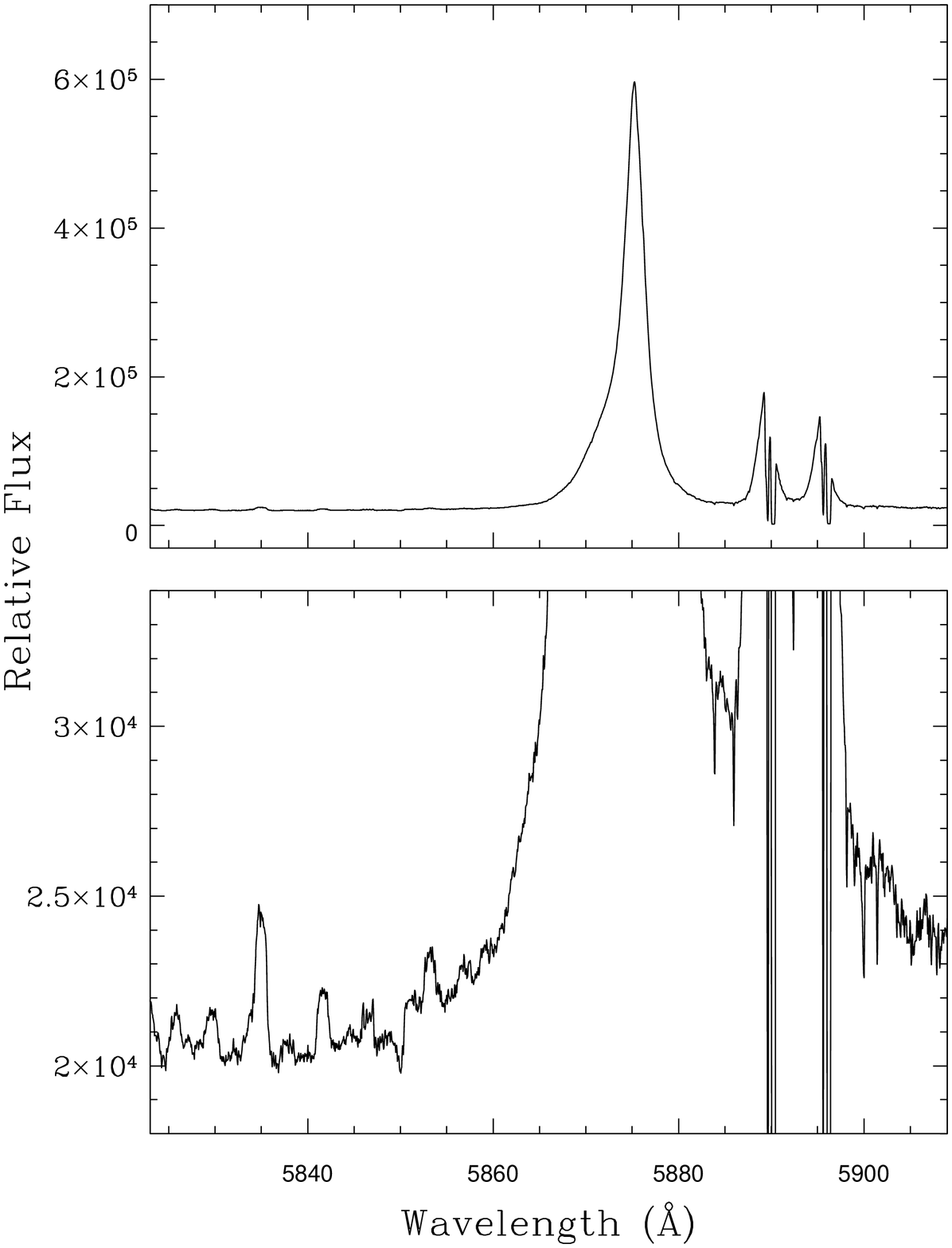,width=6.0in}

\clearpage
\epsfig{file=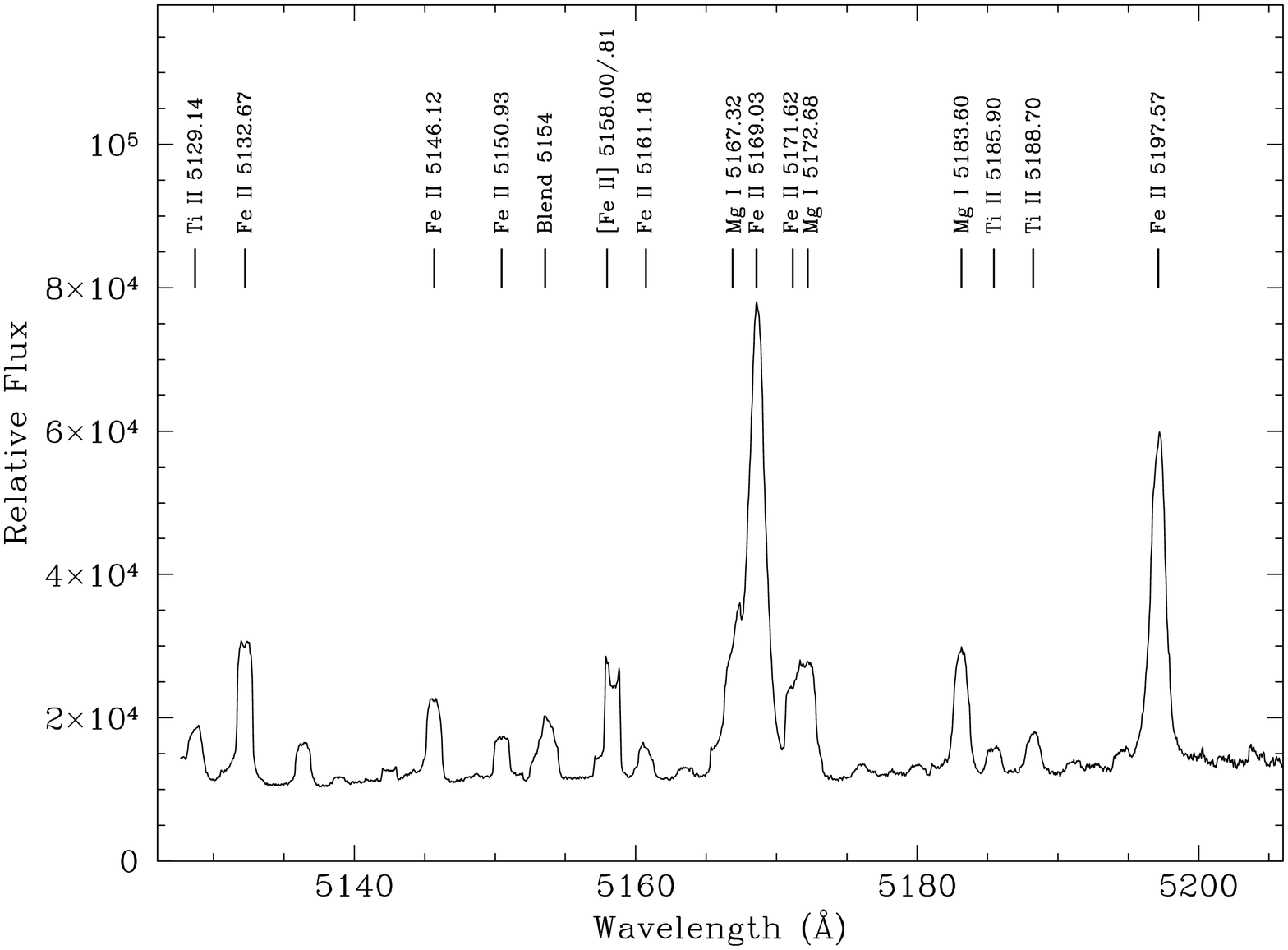,angle=-90,width=6.0in}

\clearpage
\epsfig{file=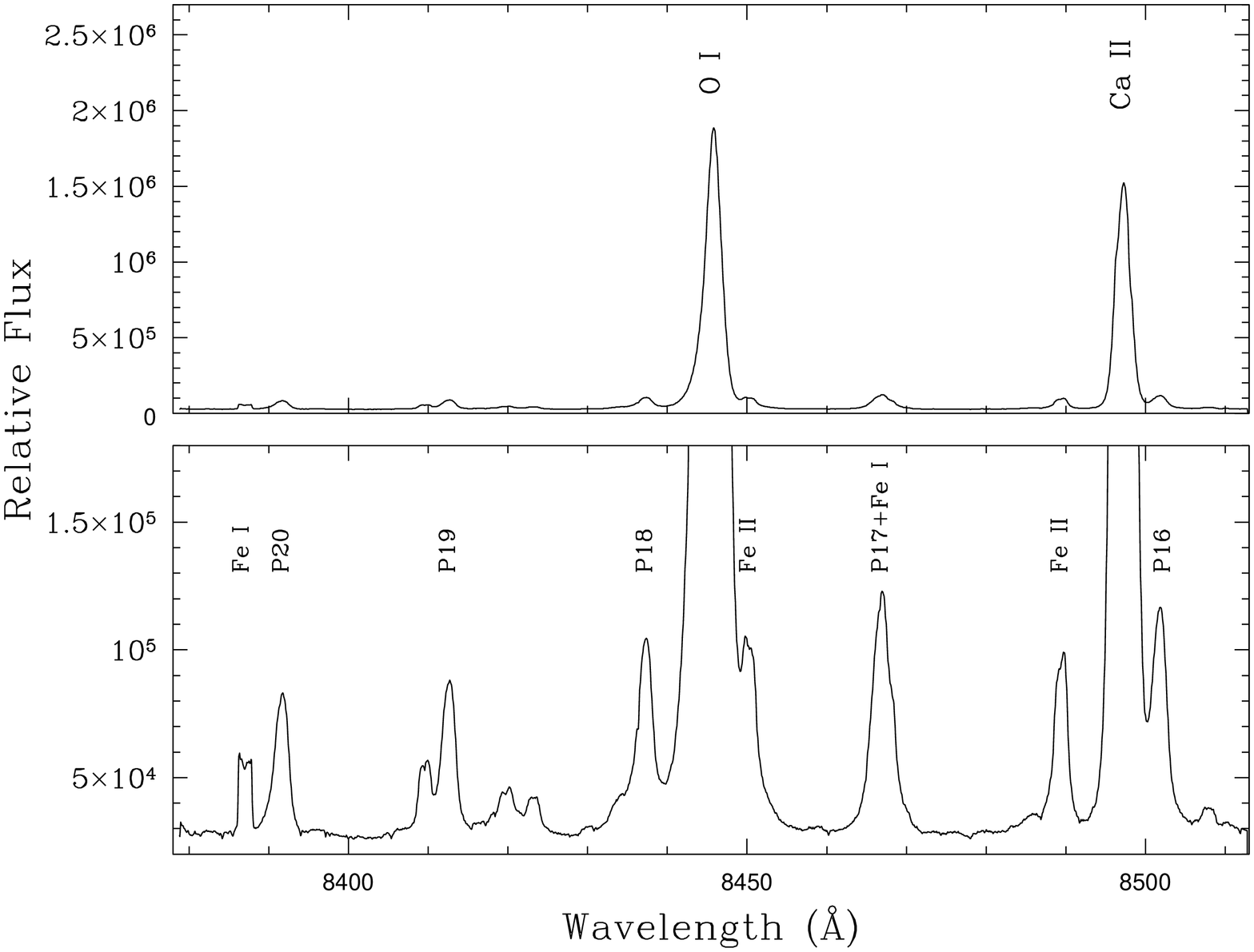,angle=-90,width=6.0in}

\clearpage
\epsfig{file=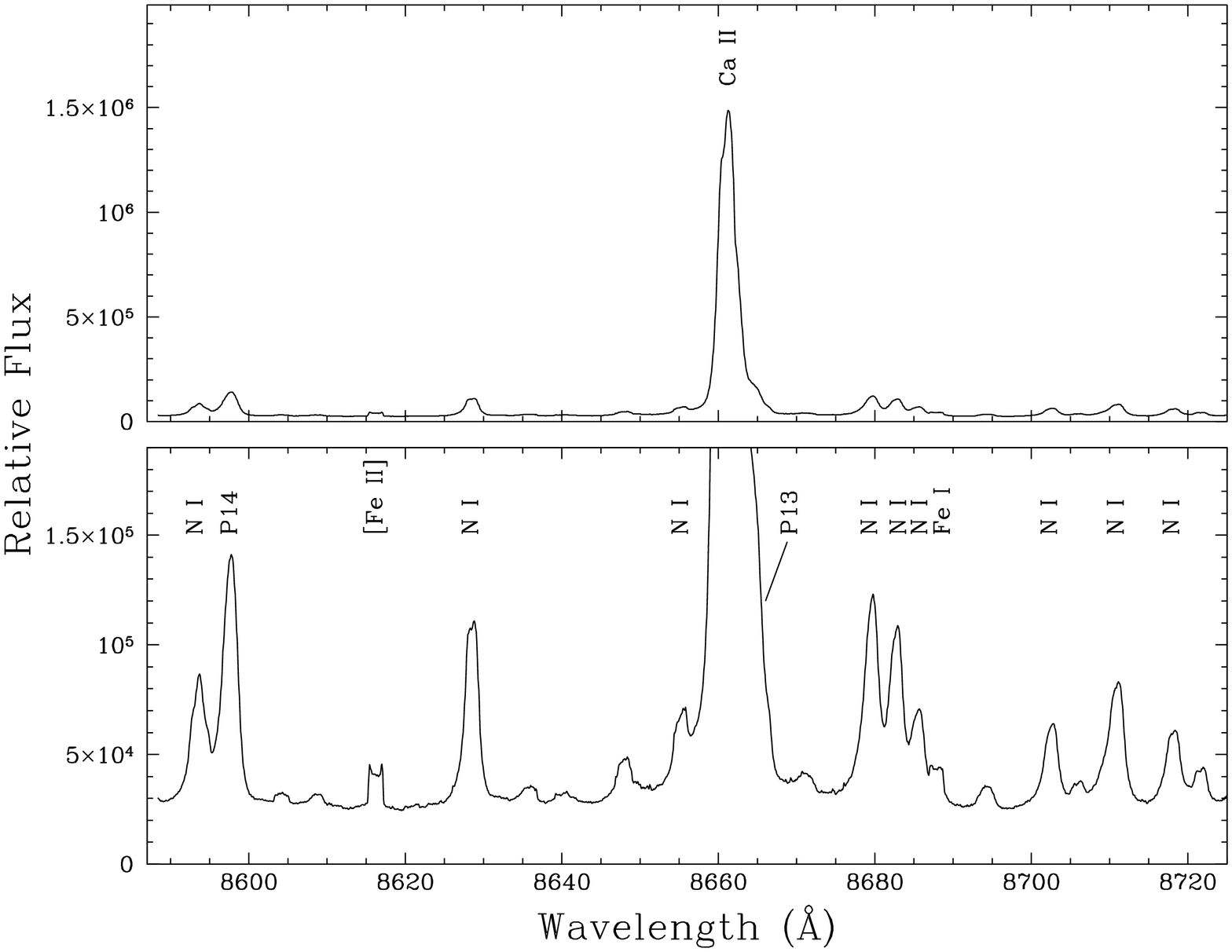,angle=-90,width=6.0in}

\clearpage
\epsfig{file=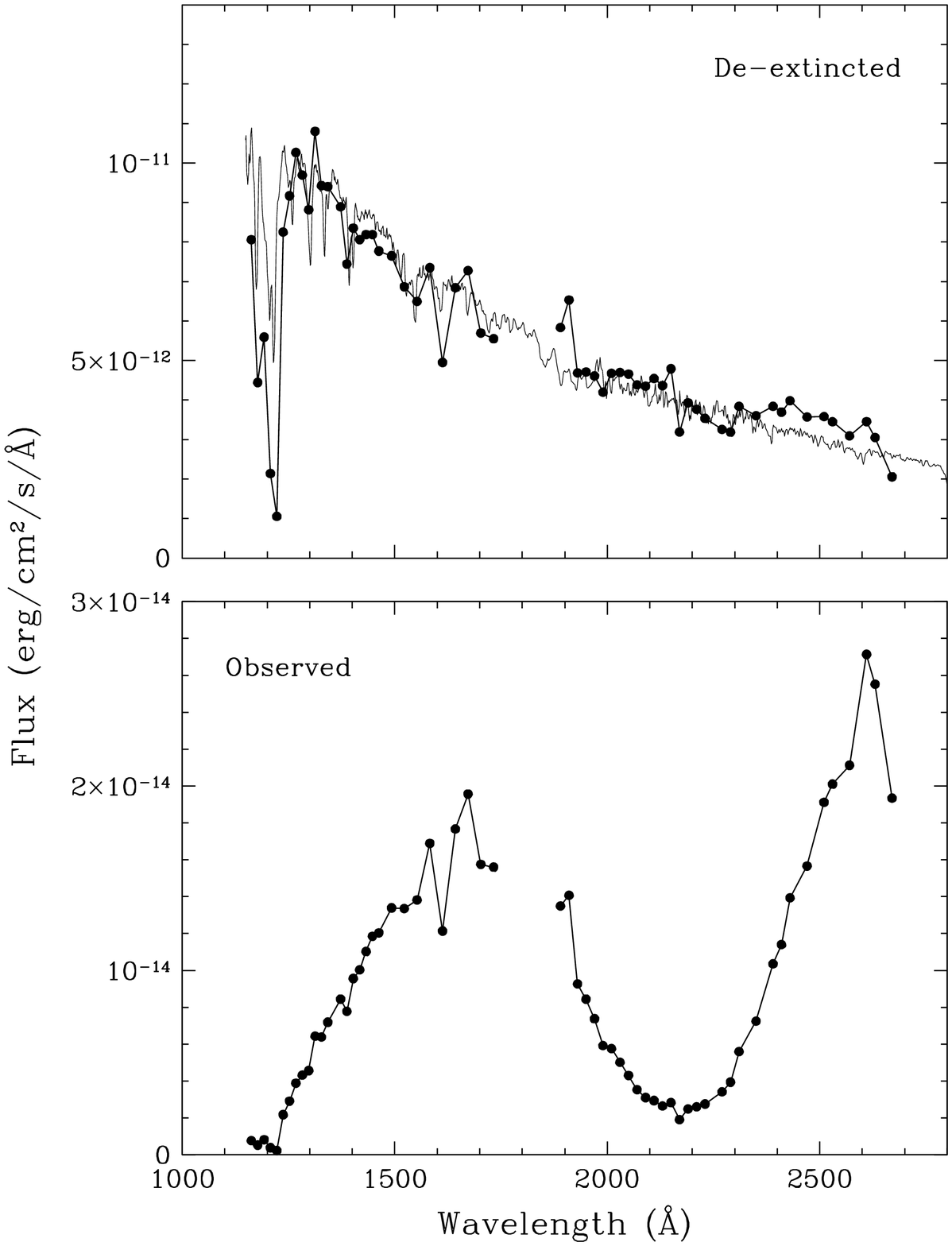,width=6.0in}

\clearpage
\epsfig{file=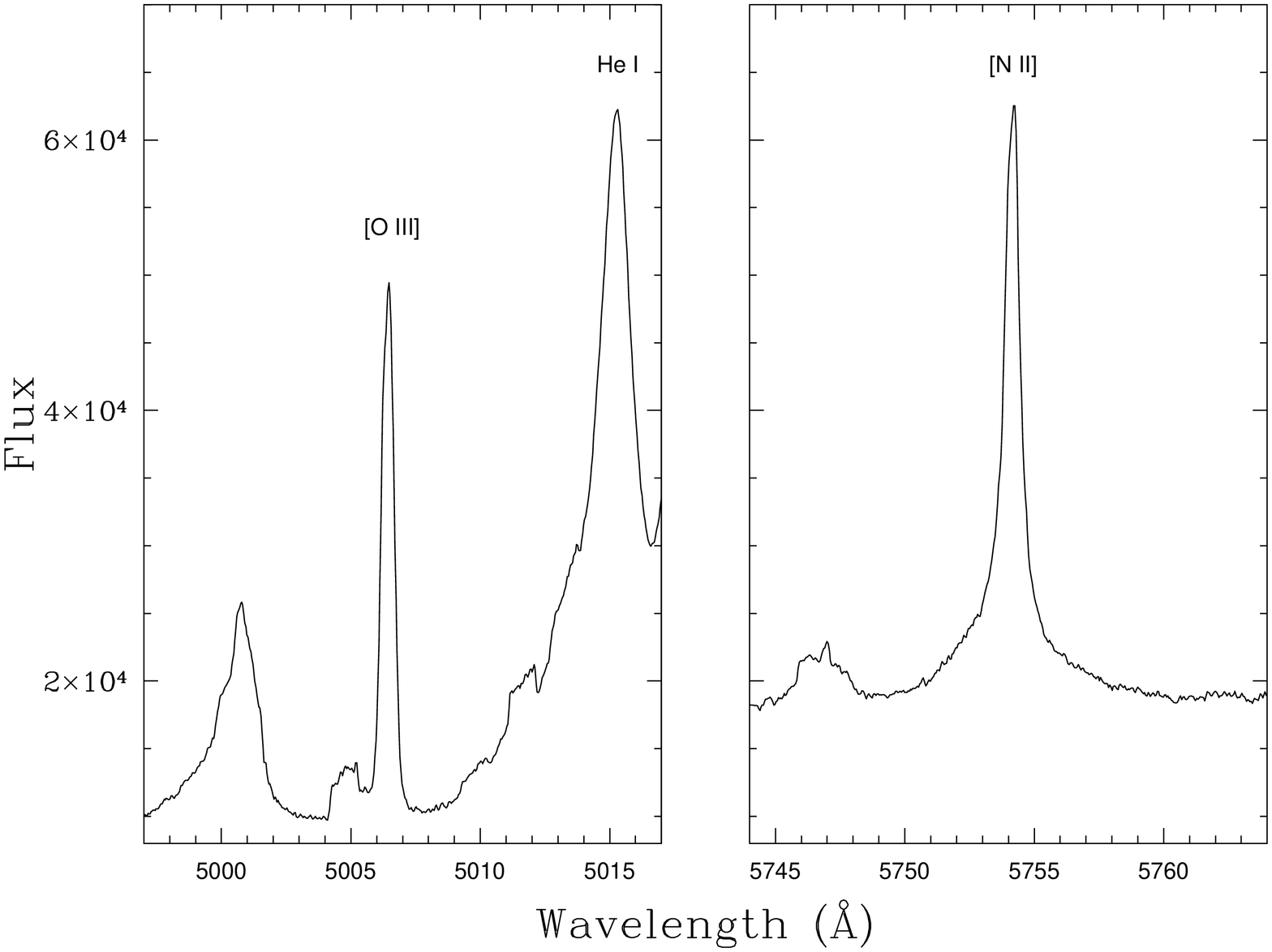,angle=-90,width=6.0in}

\clearpage
\epsfig{file=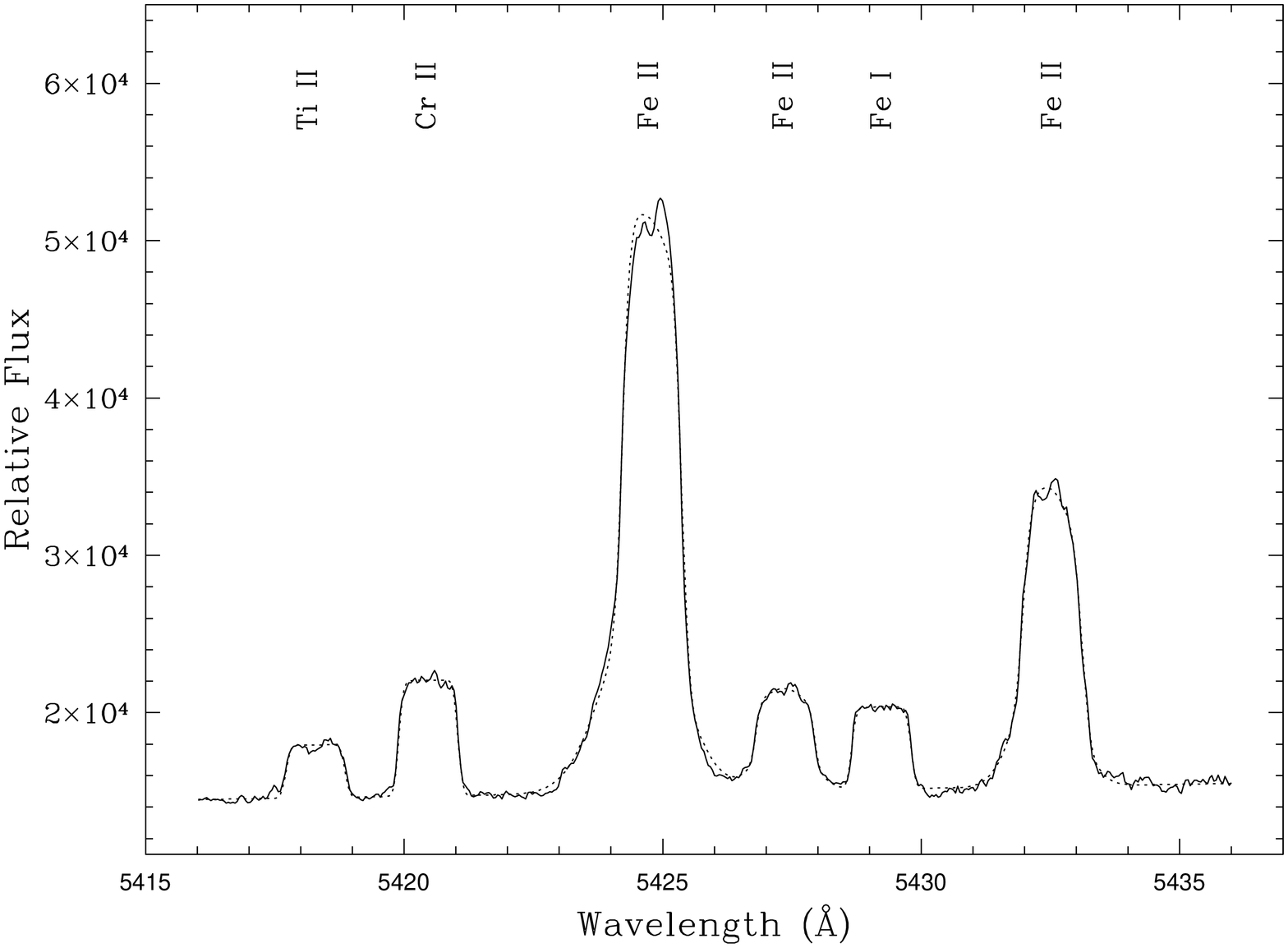,angle=-90,width=6.0in}

\clearpage
\epsfig{file=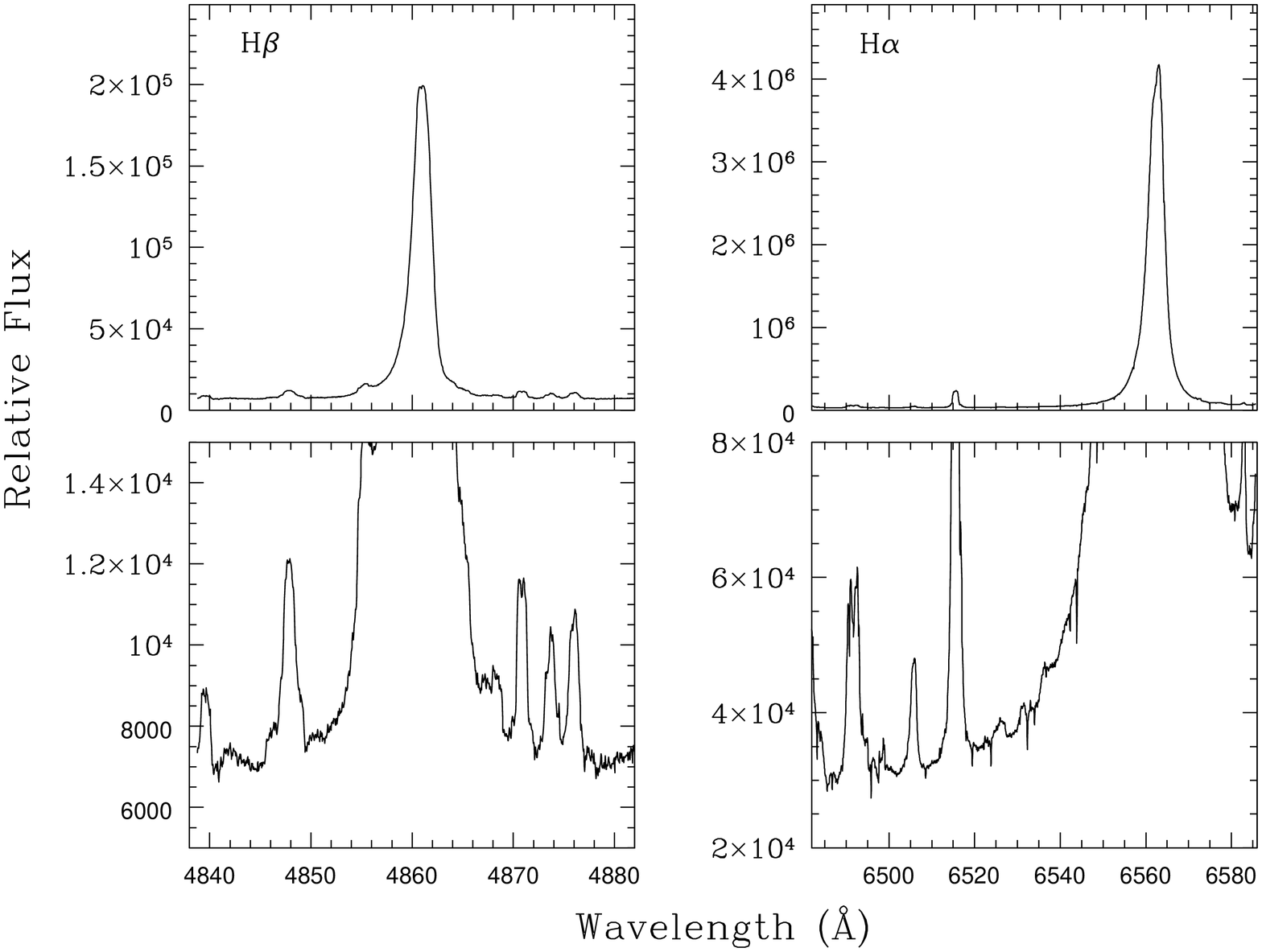,angle=-90,width=6.0in}

\clearpage
\epsfig{file=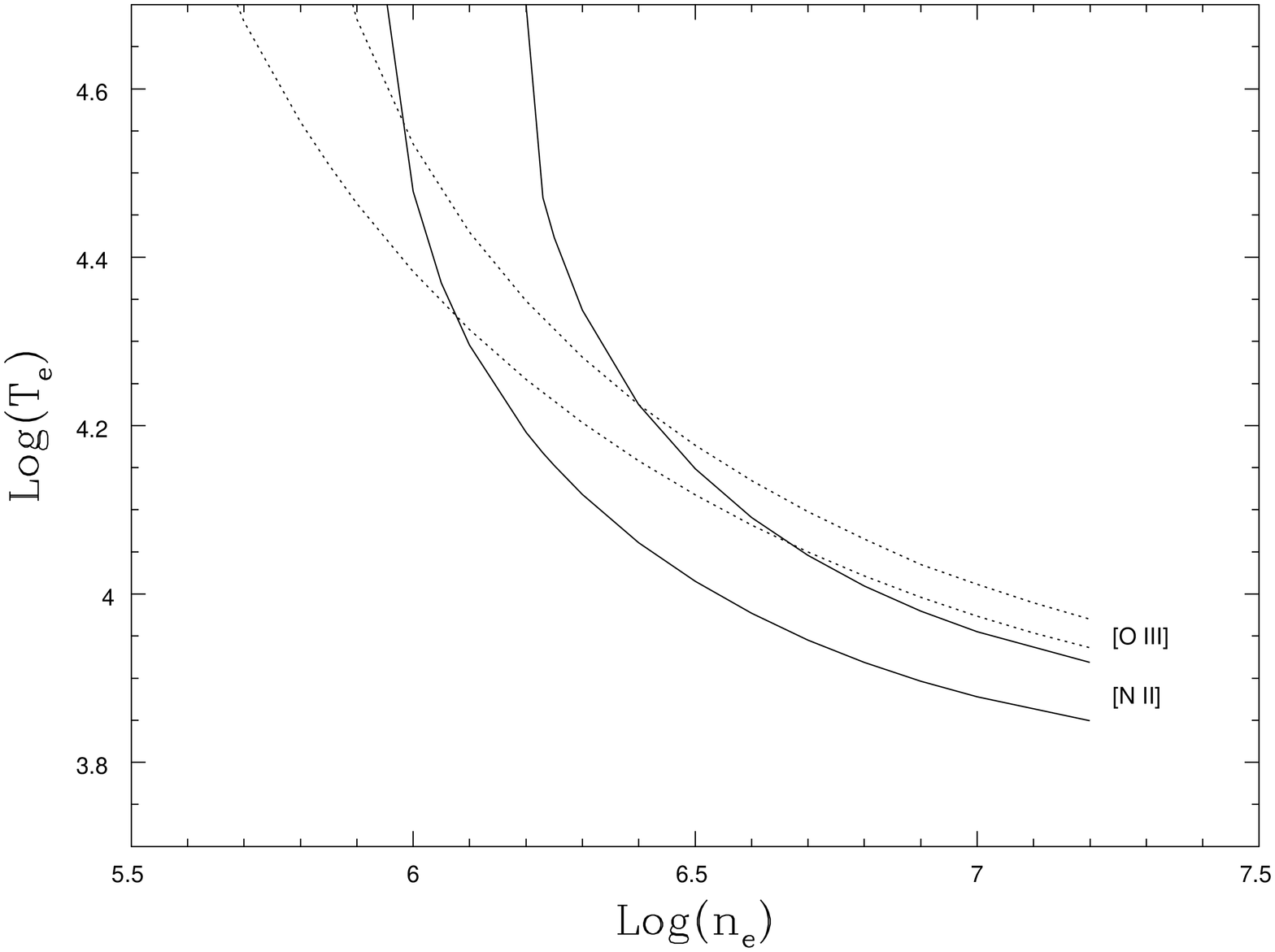,angle=-90,width=6.0in}

\clearpage
\epsfig{file=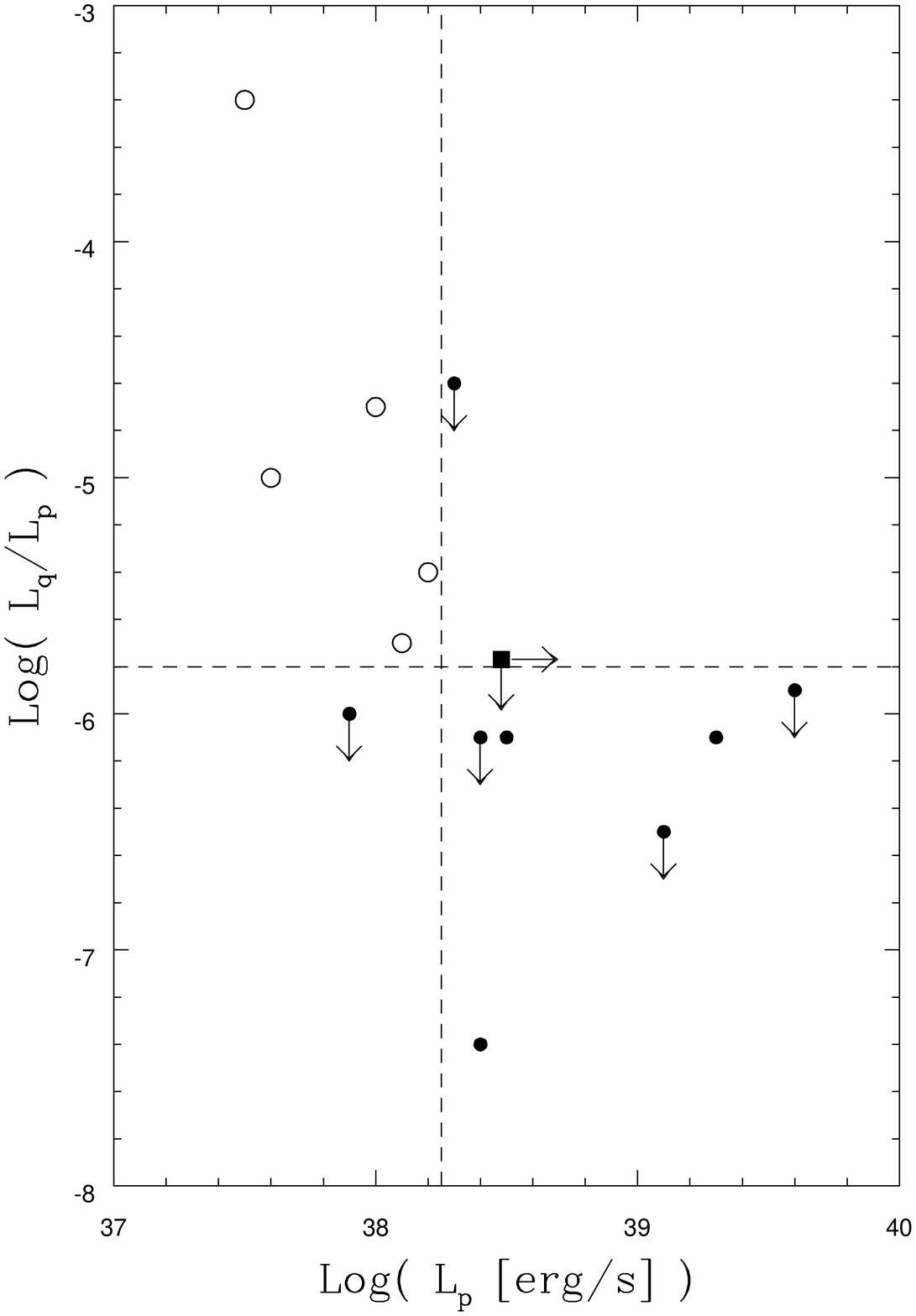,width=6.0in}


\clearpage


\begin{thebibliography}{}
\bibitem[Allen (1973)]{Alle73}
    Allen, D. A. 1973, \mnras, 161, 145

\bibitem[Barsukova et al. (1998)]{Bars98}
    Barsukova, E. A., Fabrika, S. N., Pustilnik, S. A., \&
    Ugryumov, A. V. 1998, Bull. Spec. Astrophys. Obs., 45, 147

\bibitem[Belloni et al. (1999)]{Bell99}
    Belloni, T., et al. 1999, \apj, 527, 345

\bibitem[Benjamin et al. (1999)]{Benj99}
    Benjamin, R. A., Skillman, E. D., \& Smits, D. P.
    1999, \apj, 514, 307

\bibitem[Bergner et al. (1995)]{Berg95}
    Bergner, Yu.~K. et al.~1995, \aaps, 112, 221

\bibitem[Bildsten et al. (1997)]{Bild97}
   Bildsten, L. et al. 1997, \apjs, 113, 367

\bibitem[Bjorkman (1998)]{Bjor98}
    Bjorkman, J. E. 1998, in ``B[e] Stars'', 
    ed. by A. M. Hubert \& C. Jaschek (Dordrecht: Kluwer Acad. Pub.) p. 189

\bibitem[Blitz et al. (1982)]{Blit82}
   Blitz, L., Fich, M., \& Stark, A. A. 1982, \apjs, 49, 183

\bibitem[Bondi (1952)]{Bond52}
    Bondi, H. 1952, \mnras, 112, 195

\bibitem[Burton (1988a)]{Burt88a}
    Burton, W. B. 1988a, in Galactic and Extragalactic Radio Astronomy,
    ed. by G. L. Verschuur \& K. I. Kellermann 
    (New York: Springer-Verlag) p. 295

\bibitem[Burton (1988b)]{Burt88b}
    Burton, W. B. 1988b, in The Outer Galaxy, 
    ed. by L. Blitz \& F. J. Lockman
    (Berlin: Springer-Verlag) p. 94

\bibitem[Cardelli et al. (1989)]{Card89}
    Cardelli, J. A., Geoffrey, C. C., \& Mathis, J. S.
    1989, \apj, 345, 245

\bibitem[Chan \& Fich (1995)]{Chan95}
    Chan, G., \& Fich, M. 1995, \aj, 109, 2611

\bibitem[Chen et al. (1997)]{Chen97}
    Chen, W., Shrader, C. R., \& Livio, M.
    1997, \apj, 491, 312

\bibitem[Clark et al. (2000)]{Clar00}
    Clark, J. S. et al. 2000, \aap, 356, 50

\bibitem[Corbet et al. (1997)]{Corb97}
    Corbet, R. H., Charles, P. A., Southwell, K. A., \& Smale, A. P.
    1997, \apj, 476, 833

\bibitem[Dame et al. (1987)]{Dame87}
   Dame, T. M. et al. 1987, \apj, 322, 706

\bibitem[Doty \& Leung (1994)]{Doty94}
    Doty, S. D., \& Leung, C. M. 1994, \apj, 424, 729

\bibitem[Downes (1984)]{Down84}
    Downes, R. A. 1984, \pasp, 96, 807

\bibitem[Drake \& Ulrich (1980)]{Drak80}
    Drake, S. A., \& Ulrich, R. K.
    1980, \apjs, 42, 351

\bibitem[Frank et al. (1992)]{Fran92}
    Frank, J., King, A., \& Raine, D.
    1992, Accretion Power in Astrophysics
    (Cambridge: Cambridge Univ. Press)

\bibitem[Frontera et al. (1998)]{Fron98}
    Frontera, F. et al. 1998, \aap, 339, L69

\bibitem[Garcia et al. (1998)]{Garc98}
    Garcia, M. R., McClintock, J. E., Narayan, R., \& Callanan, P. J.
    1998, in Wild Stars in the Old West, ASP Conf. Series, 137, 506

\bibitem[Grandi (1975)]{Gran75}
    Grandi, S. A. 1975, \apj, 196, 465

\bibitem[Harmon et al. (1998)]{Harm98}
    Harmon, B. A., Fishman, G. J., \& Paciesas, W. S.
    1998, \iaucirc, 6874

\bibitem[Hjellming \& Mioduszeswki (1998a)]{Hjel98a}
    Hjellming, R. M., \& Mioduszeswki, A. J. 1998a, \iaucirc, 6857

\bibitem[Hjellming \& Mioduszeswki (1998b)]{Hjel98b}
    Hjellming, R. M., \& Mioduszeswki, A. J. 1998b, \iaucirc, 6872

\bibitem[Hynes (1998)]{Hyne98}
    Hynes, Robert I. 1998, private communication

\bibitem[Ikeda et al. (2000)]{Iked00}
    Ikeda, Y., Kawabata, K. S., \& Akitaya, H.
    2000, \aap, 355, 256

\bibitem[Jaschek (1998)]{Jasc98}
    Jaschek, C. 1998, in B[e] Stars, ed. A. M. Hubert \& C. Jaschek
    (Dordrecht: Kluwer), 27

\bibitem[Jaschek \& Andrillat (2000)]{Jasc00}
    Jaschek, C., \& Andrillat, Y. 2000, in IAU Col. No. 175,
    The Be Phenomenon in Early-Type Stars, ASP Conf. Series, 214, 83

\bibitem[Kastner \& Bhatia (1995)]{Kast95}
    Kastner, S. O., \& Bhatia, A. K. 1995, \apj, 439, 346

\bibitem[Kerr et al. (1986)]{Kerr86}
    Kerr, F. J., Bowers, P. F., Kerr, M., \& Jackson, P. D.
    1986, \apjs, 66, 373

\bibitem[Kre\l owski \& Schmidt (1997)]{Krel97}
    Kre\l owski, J., \& Schmidt, M. 1997, \apj, 477, 209

\bibitem[Lamers et al. (1998)]{Lame98}
    Lamers, H. J. G. L. M., Zickgraf, F.-J., de Winter, D.,
    Houziaux, L., \& Zorec, J. 1998, \aap, 340, 117

\bibitem[Lasota (2001)]{Laso01}
    Lasota, J.-P. 2001, New Astron. Rev., 45, 449

\bibitem[Menou et al. (1999)]{Meno99}
    Menou, K. et al. 1999, \apj, 520, 276

\bibitem[Merrill (1933)]{Merr33}
    Merrill, P. W. 1933, \apj, 77, 44

\bibitem[Miroshnichenko (1995)]{Miro95}
    Miroshnichenko, A. S. 1995, A\&AT, 6, 251

\bibitem[Miroshnichenko (1998)]{Miro98}
    Miroshnichenko, A. S. 1998, in B[e] Stars, ed. A. M. Hubert \& C. Jaschek
    (Dordrecht: Kluwer), 145

\bibitem[Napiwotzki (1993)]{Napi93}
    Napiwotzki, R., 1993, Acta Astron., 43, 415

\bibitem[Narayan et al. (1997)]{Nara97}
    Narayan, R., Garcia, M. R., \& McClintock, J. E.
    1997, \apj, 478, L79

\bibitem[Neckel et al. (1980)]{Neck80}
    Neckel, T., Klare, G., \& Sarcander, M. 1980, \aaps, 42, 251

\bibitem[Netzer (1988)]{Netz88}
    Netzer, H. 1988, in IAU Coll. 94,
    Physics of Formation of Fe~II Lines Outside LTE,
    ed. R. Viotti, A. Vittone, \& M. Friedjung
    (Dordrecht: Reidel), 247

\bibitem[Orlandini et al. (2000)]{Orla00}
    Orlandini, M., et al. 2000, \aap, 356, 163

\bibitem[Orosz et al. (2001)]{Oros01}
    Orosz, J. E. et al. 2001, \apj, 555, 489

\bibitem[Osterbrock et al (1997)]{Oste97}
    Osterbrock, D. E., Fulbright, J. P., \& Bida, T. A.
    1997, \pasp, 109, 614

\bibitem[Oudmaijer et al. (1998)]{Oudm98}
    Oudmaijer, R. D., Proga, D., Drew, J. E., \& de Winter, D.
    1998, \mnras, 300, 170

\bibitem[Pacheco (1998)]{Pach98}
    Pacheco, J. A. De Freitas 1998, 
    in B[e] Stars, ed. A. M. Hubert \& C. Jaschek
    (Dordrecht: Kluwer), 221

\bibitem[Paciesas \& Fishman (1998)]{Paci98}
    Paciesas, W. S., \& Fishman, G. J. 1998, \iaucirc, 6856

\bibitem[Predehl \& Schmitt (1995)]{Pred95}
    Predehl, P., \& Schmitt, J. H. M. M. 1995, \aap, 293, 889

\bibitem[Revnivtsev et al. (1999)]{Revn99}
    Revnivtsev, M., Emelyanov, A., \& Borozdin, K.
    1999, Astron. Lett., 25, 296

\bibitem[Robinson et al. (1998)]{Robi98}
    Robinson, E. L., Welsh, W. F., Adams, M. T., \& Cornell, M.~E.
    1998, \iaucirc, 6862

\bibitem[Rybicki \& Lightman (1979)]{Rybi79}
    Rybicki, G. P., \& Lightman, A. P.
    1979, Radiative Processes in Astrophysics 
    (New York: Wiley).

\bibitem[Sciortino et al. (1990)]{Scio90}
    Sciortino, S. et al. 1990, \apj, 361, 621

\bibitem[Shaw \& Dufour (1995)]{Shaw95}
    Shaw, R. A., \& Dufour, R. J. 1995, \pasp, 107, 896

\bibitem[Smith \& Remillard (1998)]{Smit98}
    Smith, D. \& Remillard, R. 1998, \iaucirc, 6855

\bibitem[Taylor \& Cordes (1993)]{Tayl93}
    Taylor, J. H., \& Cordes, J. M. 1993, \apj, 411, 674

\bibitem[Tobin (1985)]{Tobi85}
    Tobin, W. 1985, \aap, 142, 189

\bibitem[Traub (1999)]{Trau99}
    Traub, W. A. 1999, private communication.

\bibitem[Tull et al. (1995)]{Tull95}
    Tull, R. G., MacQueen, P. J., Sneden, C., \& Lambert, D. L.
    1995, \pasp, 107, 251

\bibitem[Ueda et al. (1998)]{Ueda98}
    Ueda, Y., Ishida, M., Inoue, H., Dotani, T., Greiner, J.
    \& Lewin, W. H. G.
    1998, \apj, 508, L167

\bibitem[Verner et al. (1999)]{Vern99}
    Verner, E. M., Verner, D. A., Korista, K. T., Ferguson, J. W.,
    Hamann, F., \& Ferland, G. J.
    1999, \apjs, 120, 101

\bibitem[Wagner \& Starrfield (1998)]{Wagn98}
    Wagner, R. M., \& Starrfield, S. G. 1998, \iaucirc, 6857

\bibitem[Zickgraf et al. (1985)]{Zick85}
    Zickgraf, F.-J., Wolf, B., Stahl, O., Leitherer, C., \& Klare, G.
    1985, \aap, 143, 421

\bibitem[Zickgraf et al. (1986)]{Zick86}
    Zickgraf, F.-J., Wolf, B., Stahl, O., Leitherer, C., \& Appenzeller, I.
    1986, \aap, 163, 119

\bibitem[Zickgraf (1998)]{Zick98}
    Zickgraf, F.-J. 1998, in ``B[e] Stars'', 
    ed. by A. M. Hubert \& C. Jaschek (Dordrecht: Kluwer Acad. Pub.) p. 1

\bibitem[Zorec (1998)]{Zore98}
    Zorec, J. 1998, in B[e] Stars, ed. A. M. Hubert \& C. Jaschek
    (Dordrecht: Kluwer), 27
\end{thebibliography}
\end{document}